\documentclass[twocolumn,amssymb,nobibnotes, aps, pra,showpacs,superscriptaddress]{revtex4-1}

\usepackage{amsmath}
\usepackage{verbatim}
\usepackage{epsfig}
\usepackage{subfigure}
\usepackage[breaklinks]{hyperref}
\usepackage{graphicx}  % needed for figures
\usepackage{dcolumn}   % needed for some tables
\usepackage{bm}        % for math
\usepackage{amssymb}   % for math
\usepackage{color}
\usepackage{adjustbox}
\newcommand{\ket}[1]{\left|{#1}\right\rangle}

\newcommand{\aver}[1]{\left\langle{#1}\right\rangle}

\newcommand{\modu}[1]{\left|{#1}\right|}

\newcommand{\bea}{\begin{eqnarray}}
\newcommand{\eea}{\end{eqnarray}}
\begin{document}

\title{Nonlinear dynamics of superpostion of wavepackets}%
\author{S. Kannan}
\affiliation{Department of Physics, Indian Institute of Space Science and Technology, Thiruvananthapuram, 695 547, India.}
\author{M. Rohith}%
%\email{rohith.manayil@gmail.com}
\affiliation{Center for Theoretical Physics of Complex Systems, Institute for Basic Science (IBS), Daejeon 34126, Republic of Korea}
\author{C. Sudheesh}
\email{sudheesh@iist.ac.in}
\affiliation{Department of Physics, Indian Institute of Space Science and Technology, Thiruvananthapuram, 695 547, India.}
\date{\today}
\begin{abstract}
We study nonlinear dynamics of superposition of quantum  wavepackets in various systems such as Kerr medium, Morse oscillator and  bosonic
Josephson junction. The prime  reason  behind this study is to find out how the superposition of states influence  the dynamics of quantum systems.
 We consider the superposition states which are potential candidates for quantum computing and quantum communication and so it is most necessary that we study the dynamics for their proper understanding and usage.
 Methods in nonlinear time
series analysis such as first return time distribution, recurrence
plot and  Lyapunov exponent are used for the qualification and
quantification of dynamics.
We found that there is a vast change in the dynamics of quantum systems when we consider the superposition of wave packets. These changes are observed in various kinds of dynamics such as periodic, quasi-periodic, ergodic, and chaotic dynamics.

\end{abstract}
\maketitle
%\tableofcontents
\section{Introduction}\label{sec1}
Quantum superposition is one of the most fundamental features of quantum mechanics \cite{dirac1981}  with which one can explain the quantum effects arising from the interference of quantum amplitudes.
In classical physics, it is possible to have a superposition of fields which will give rise to a new field  but  the quantum mechanical concept  of probability to occur the individual states is not feasible \cite{da2013paraconsistent}. The properties of these quantum superposition states can be used in various applications  of quantum information theory such as quantum communication, quantum teleportation, quantum cryptography, quantum cloning \cite{feix2015,bennett1992,milburn1999,ralph2003,cerf2000} %put [2,3,4]
etc. It is a well-established fact that various non-classical
effects such as squeezing, higher-order squeezing, sub-Poissonian statistics and oscillations of the photon number distribution are exhibited by superposition of coherent states \cite{buvzek1992superpositions,gerry1993non} in contrast to ordinary coherent states. % Give reference to superposition state squeezing
Various theoretical \cite{yurke1986,miranowicz1990,paprzycka1992,tara1993} and experimental methods \cite{monroe1996schrodinger,ourjoumtsev2007generation,gao2010experimental} are also available for the production of superposition states. The  experimental observation of Schr\"{o}dinger cat states, which is a superposition state, using the single-photon Kerr effect  has opened new directions in continuous variable quantum communication \cite{kirchmair2013}. %put refe 18 here.

On the other hand, extensive studies have been carried out on the dynamics of quantum systems but less has been done for the dynamics of superposition states.
Ergodicity in quantum systems has received much attention after the quantum ergodic theory proposed by von Neumann in as early 1929  \cite{von1929}. % more about this.
Later, Peres  gave the newest definition of quantum ergodicity  as the time average of any quantum operator equal to its average of microcanonical ensemble \cite{peres1984}. If the motion evolves to exponentially  separated trajectories even for nearly identical initial conditions, such types are referred to as chaotic. Chaos is a type of motion that lies between the regular deterministic trajectories arising from solutions of integrable equations and a state of noise or stochastic behaviour characterized by complete randomness \cite{goldstein2014}. Nonlinear dynamics  of quantum systems have been of special interest and have been studied by many \cite{scott2001quantum,maletinsky2007nonlinear,bettelheim2007nonlinear,wang2014nonlinear}. Various methods are available to study the nonlinear dynamics of quantum systems such as  random matrix theory \cite{mehta2004random,andreev1996quantum}, recurrence time distributions and recurrence quantification analysis \cite{eckmann1987,marwan2007} and Lyapunov spectra \cite{rosenstein1993,  kantz1994}. In the literature,   expectation values of certain dynamical variables are  considered as time series  to study quantum dynamics of various systems \cite{sudheesh2009,sudheesh2010,shankar2014dynamics,pradeep2019dynamics}.
There are a few studies addressing the dynamics of superposition of wave packets, for example, fractional revivals of superposed wave packets in a nonlinear Hamiltonian \cite{rohith2014fractional,rohith2015tomogram}. However, qualitatively different, such as quasi-periodic, ergodic, and chaotic behavior in the dynamics of a superposition of quantum wave packets are not reported.
In this review  paper, we would like to study in detail  the dynamics of superposition of quantum wavepackets and  investigate how the superposition alter  the dynamics of quantum systems.
We use time series generated from  expectation values  for  studying  the dynamics of superposition states  to show the differences between superposition states and non-superposition states in terms of  periodic, ergodic and chaotic dynamics.

This paper is organized as follows. In Sec. \ref{sec2}, we study and analyze the
periodic properties of expectation values of initial   superposition states  for two different quantum systems which are governed by nonlinear Hamiltonians.  In Sec.
\ref{sec3}, we find the first return time distribution, recurrence
plot and Lyapunov exponent  using  time series data of expectation values for different
quantum states and its superposition states.  The chaotic and ergodic properties of the systems are analyzed in this section.  In Sec.
\ref{sec4}, we summarize the main results of the paper.

\section{Periodicity of expectation values}\label{sec2}
The dynamics of a normalized quantum system $ \ket{\psi(0)} $ is said to be periodic if the autocorrelation function $ \modu{\langle{\psi(t)}|{\psi(0)}\rangle}^{2} $ becomes unity, where $ \ket{\psi(t)} $ is the time evolved state. When a quantum system is periodic, all expectation values of the system attains its initial value periodically. Various quantum systems showing periodic dynamics can be seen in the literature \cite{robinett2004}. Harmonic oscillator, infinite well etc., are popular systems showing periodic dynamics. We will be studying similar periodicity in the time evolution of  quantum systems governed by nonlinear Hamiltonians.
% general statement periodic dynamics of expectation value	various examples in literature to show the perioicity.
In this section, we show how the period of expectation values of certain quantum variables changes with respect to the initial states which are superposition of quantum states. For this purpose, we consider the dynamics of superposition states  in  Kerr medium and  Morse oscillator.
\begin{figure*}[t]
	\centering
	\includegraphics[scale=0.6]{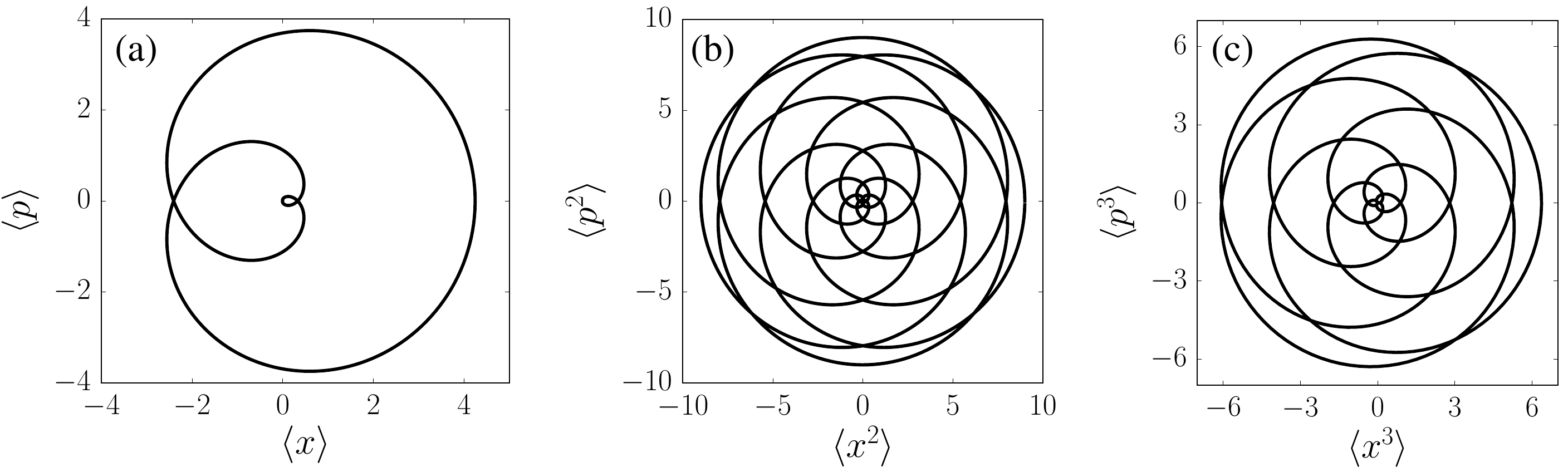}
	\caption{Plots showing the revival of (a) coherent state ($l=1$) and its superposition states with (b)  $ l=2$ and (c)  ($l=3$). The expectation values are plotted from $t=0$ to (a) $t= \frac{\pi}{\chi} $,  (b) $ \frac{\pi}{\chi} $ and (c) $ \frac{\pi}{3\chi} $. These closed curves indicates that  these states are having periodic dynamics with different time period.}\label{fig1}
\end{figure*}
\subsection{Dynamics of superposition states  in a Kerr medium}
	Consider the dynamics governed by a nonlinear Hamiltonian which is the effective Hamiltonian for the propagation of coherent field in a Kerr medium \cite{milburn1986quantum,kitagawa1986number}
\begin{equation} \label{kerr}
H=\hslash\chi {\hat{a}^{\dag 2}} \hat{a}^{2}=\hslash\chi {N}({N}-1),
\end{equation}
where $\hat{a}$ and $ \hat{a}^{\dag}$ are annihilation and creation operators. The operator ${N} = {a}^{\dag}{a}$ is the number operator whose eigenstates are the Fock state $\ket{n}$, where $n=0,\,1,\,2\,\dots\infty$ and $\chi$ is the nonlinear susceptibility of the medium.
Consider a general wavepacket $ \ket{\psi(0)} $ which can be expanded in the Fock basis as 					\begin{equation}
\ket{\psi(0)}=\sum_{n=0}^{\infty}C_{n}\ket{n},
\end{equation}
where $ C_{n} $ are the Fock state expansion coefficients. Using the unitary time evolution opetator $e^{-iHt/\hslash}$,  we can obtain the state at time $t$:
\begin{equation}
\ket{\psi(t)}=e^{-iHt/\hslash}\ket{\psi(0)}.
\end{equation}

Let a coherent state  $ \ket{\alpha(0)} $ be the initial wave packet which is the eigenstate of the annihilation operator $ \hat{a} $ with eigenvalue $ \alpha $.
The  state at time $t$ can be expressed in the Fock basis as
	\begin{equation}
	\ket{\alpha(t)}=e^{-{\modu{\alpha}^{2}}/{2}}\sum_{n=0}^{\infty}e^{-i\chi n(n-1)t}\frac{\alpha^{n}}{\sqrt{n!}}\ket{n}.
	\end{equation}
At time $t= \pi/\chi$ and its integer multiple instants, the system becomes periodic. In other words, at these instants the fidelity  $ \modu{\langle{\alpha(t)}|{\alpha(0)}\rangle}^{2}$ becomes unity.  This result was already appeared in \cite{tara1993}. From now on, we use $T_{per}$  to denote the time period of systems which are having periodicity. Our interest is to find how the time period $T_{per}=\pi/\chi$ changes when we consider initial states which are superpositions of coherent states. For this purpose, we consider a general superposition of $ \ell $ coherent states  \cite{napoli1999}

\begin{equation}\label{key1}
\ket{\Psi_\ell(0)}=N_\ell\sum_{j=0}^{\ell-1}\ket{\alpha\varepsilon_{j}^{(\ell)}}.
\end{equation}
where   $N_\ell$ is the normalization constant and
 \begin{equation}
\varepsilon_{j}^{(\ell)}\label{epsilon} \equiv e^{i{2\pi}j/{\ell}}.
\end{equation}
The superposition state $ \ket{\Psi_\ell(0)} $ can be expanded in the Fock basis as
\begin{equation}
\ket{\Psi_\ell(0)}=N_\ell\sum_{n=0}^{\infty}\frac{\alpha^{\ell n}}{\sqrt{(\ell
n)!}}\ell\ket{\ell n}.
\end{equation}

To derive the above expression, we have  used the well-known identity
\begin{equation}
\sum_{j=0}^{\ell-1}(\varepsilon_{j}^{(\ell)})^{n}=\ell\delta_{[\frac{n}{\ell}],\frac{n}{l}},
\end{equation}
where $ \delta $ is the Kronecker delta  and $[\,\,\,]$  denotes the
greatest integer function. The  state at any time $t$, using the Kerr Hamiltonian given in Eq. (\ref{kerr}), is
\begin{equation}
\ket{\Psi_l(t)}=\textit{N}_l\sum_{n=0}^{\infty}e^{-i\chi \ell n(\ell
n-1)t}\frac{\alpha^{\ell n}}{\sqrt{\ell n!}}\ell\ket{\ell n}.
\end{equation}

The above equation is for a general superposition of $\ell$ coherent states.
Now we specifically  look at the case $\ell=2$.
With $l=2$ in Eq. (\ref{key1}), we get a superposition of two coherent states $\ket{\alpha}$ and $\ket{-\alpha}$ which is known as the even coherent state \cite{dodonov1974}.

The time evolution of an initial even coherent state gives
	\begin{equation}
	\ket{\Psi_2(t)}=N_2\sum_{n=0}^{\infty}e^{-i\chi 2n(2n-1)t}\frac{\alpha^{2n}}{\sqrt{2n!}}2\ket{2n}.
	\end{equation}
It is evident that this state also has the same periodicity of $\pi/\chi$ obtained for initial coherent state.
\begin{figure*}[t]
	\centering
	\includegraphics[scale=0.6]{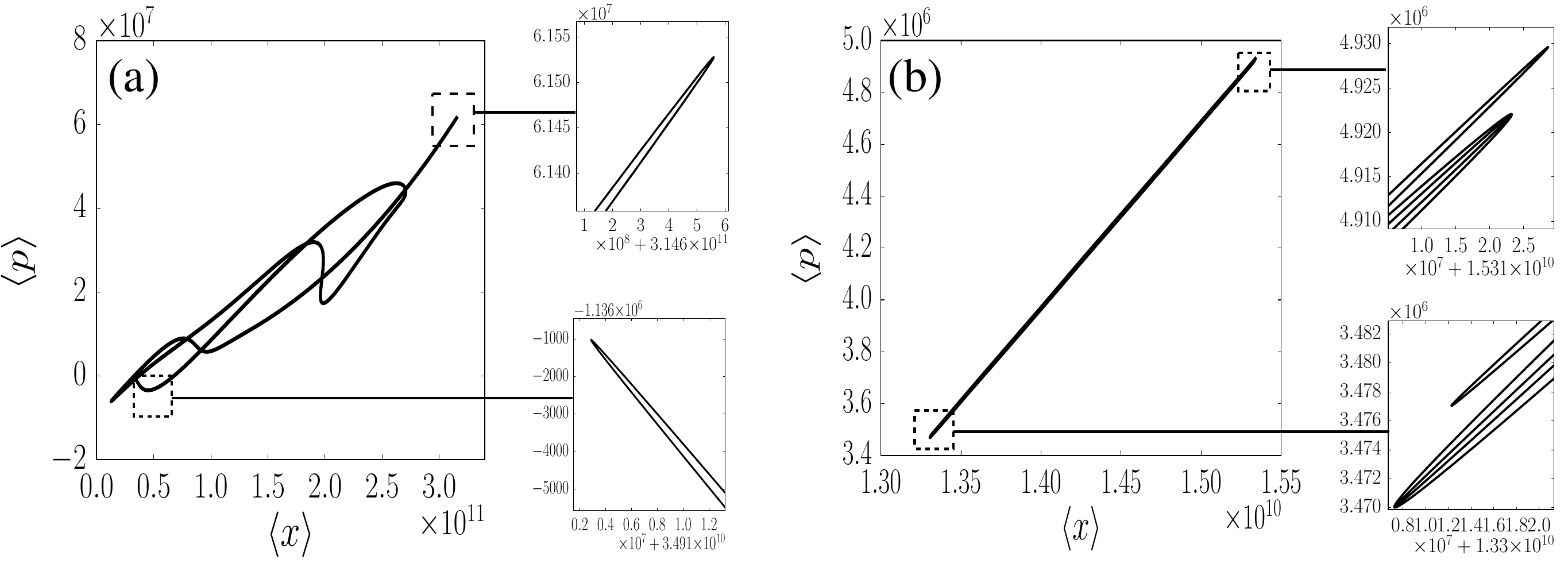}
  \caption{Plots showing the periodic dynamics  of  (a) coherent state ($l=1$) and (b) its
superposition state ($l=2 $). The expectation values are plotted from $t=0$ to
(a) $t=T_{per}$, and (b) $T_{per}/4$. The closed curves
indicates that  the variables $\aver{x}$ and $\aver{p}$ return to its initial values at   these
 instances.}\label{fig2}

\end{figure*}
However, when we increase the value of $\ell$, the periodicity depends on the value of $\ell$. The following results can be obtained from Eq. (11):   If $ \ell $ is an even number then  $ T_{per}={2\pi}/{\chi\ell} $. If it is odd then $ T_{per}={\pi}/{\chi\ell}$.

Figure \ref{fig1} shows the above results for certain values of $\ell$ using the expectation values and higher moments  of the dynamical quantities $ x $ and $ p $  where
\begin{eqnarray}
x&=&\dfrac{\hat{a}+\hat{a}^{\dagger}}{\sqrt{2}} \qquad \text{and}\\
p&=&\dfrac{\hat{a}-\hat{a}^{\dagger}}{i\sqrt{2}}.
\end{eqnarray}
Figure \ref{fig1}(a) shows the plot of  $ \langle x \rangle$ versus $\langle p \rangle$  for the case $\ell=1 $. Figure \ref{fig1}(b) is plotted between $\langle x^{2} \rangle $ and $ \langle p^{2} \rangle $ for the case  $ \ell=2 $ state and Fig. \ref{fig1}(c) is between $ \langle x^{3} \rangle $ and $ \langle p^{3} \rangle $ for $ \ell=3 $. The closed curve in all the plots is an indicator of periodic dynamics for corresponding  quantum states.  We have shown that  periodicity of the motion changes when we  change the  initial state to a  superposition state. In the next session, we consider the dynamics of  Morse oscillator system which also illustrate similar results.

\subsection{Periodic dynamics of Morse oscillator}%%Perolomov
The Morse oscillator is a model for a particle in a one-dimensional
anharmonic potential energy surface with a dissociative limit at
infinite displacement\cite{morse1929}. The Morse potential describing
the vibrational motion of a diatomic molecule can be expressed as
\begin{equation}\label{eq17}
V(x)=D(e^{-2\beta x}-2e^{-\beta x}),
\end{equation}
where $D$ is the dissociation energy, $\beta$ is a range parameter and
$x$ gives the relative distance from the equilibrium position. The
eigenfunctions of the Morse potential for the reduced one body system
can be written as \begin{equation}\label{eq18}
\Psi_{n}^{\lambda}(\xi)=\text{N}e^{-\xi/2}\xi^{s/2}L_{n}^{s}(\xi),
\end{equation} where $ \xi=2\lambda$$ e^{-\beta x} $, $ 0<\xi<\infty $
and $ n=0,1,\dots,[\lambda-1/2] $, with $ [\text{ }] $ being the
greatest integer function. Here $ \lambda $ and $ s $ are the
potential and energy dependent parameters:
\begin{equation}\label{eq19}
\lambda=\sqrt{\frac{2\mu Dr_{0}^{2}}{\beta^{2}\hslash^{2}}}, \text{ }
s=\sqrt{\frac{-8\mu r_{0}^{2}E}{\beta^{2}\hslash^{2}}}
\end{equation}
where $ \mu $ is the reduced mass of the system and $r_{0} $ is
the equilibrium position. In Eq. (\ref{eq18})  $L_{n}^{s} $ is the
associated Laguerre polynomial and N  the normalization constant given
by\begin{equation}\label{eq20}
\text{N}=\left[\dfrac{\beta(2\lambda-2n-1)\Gamma(n+1)}{\Gamma(2\lambda-n)r_{0}}\right]^{1/2}.
\end{equation}

 Using the annihilation and creation operators for the Morse
Hamiltonian \cite{dong2002}, a displacement operator is defined
\cite{ghosh2006} which acts on the highest bound state $
(\Psi_{n'}^{\lambda}(\xi)) $ to produce a Perelomov coherent state.
The general expression for the coherent state is given as
\begin{equation}\label{eq21}
\chi(\xi)=\sum_{n=0}^{n'}d_{n}\Psi_{n}^{\lambda}(\xi).
\end{equation}
where ${n'}$ corresponds to the highest bound state and
\begin{equation}\label{eq22}
d_{n}=\frac{(-\alpha)^{n'-n}}{(n'-n)!}\sqrt{\frac{n'!\Gamma(2\lambda-n)}{n!\Gamma(2\lambda-n')}}.
\end{equation}
The distribution of $d_{n} $ or the value of $ \alpha $ decides the
localization of the coherent state. To study the dynamics under Morse
potential, let us look at the time evolution of this wave function
\begin{equation}\label{eq23}
\chi(\xi,t)=\sum_{n=0}^{n'}d_{n}\Psi_{n}^{\lambda}(\xi)e^{{-iE_{n}t}/{\hslash}}.
\end{equation}
For the Morse oscillator \cite{strekalov2016},
\begin{equation}\label{eq24}
E_{n}=E_{0}+\hslash(\omega n-x_{e}\omega_{e} n^{2})
\end{equation}
where the harmonic frequency is given by
\begin{equation}\label{eq25}
\omega_{e} = \dfrac{\hslash\beta^{2}}{\mu}\left(N_{n}+\dfrac{1}{2}\right)=
\sqrt{\frac{2D\beta^{2}}{\mu}}.
\end{equation} $ N_{n} $ can be identified as $(\lambda-1/2) $ and the
frequency $ \omega $ differs from $ \omega_{e} $ by the anharmonicity
constant $x_{e}$
\begin{equation}\label{eq26}
\omega=\omega_{e}(1-x_{e}),
\end{equation} with \begin{equation}\label{eq27}
x_{e}=\dfrac{1}{2N_{n}+1}.
\end{equation} The periodic dynamics of the system is calculated using
the spatially averaged autocorrelation function $ A(t) $,
\begin{equation}\label{eq28}
A(t)=\int_{-\infty}^{\infty}\chi^{*}(\xi,0)\chi(\xi,t)d\xi=\sum_{n=0}^{n'}\modu{c_{n}}^{2}e^{-iE_{n}t/\hslash}.
\end{equation}
With the assumption that the zero point energy is the reference zero
of energy, Eq. \ref{eq28} can be rewritten as
\begin{equation}\label{eq29}
A(t)=\sum_{n=0}^{n'}\modu{c_{n}}^{2}\text{exp}\left(-in\omega t
+in^{2}x_{e}\omega_{e}t\right),
\end{equation}
where $ c_{n} $ are the weighting coefficients. Periodicity of the
system occurs at those instances when $ A(t)=A(0) $ and this
periodicity time, $ T_{per} $ can be calculated by equating the
exponential to 1:
\begin{equation}\label{eq30}
\text{exp}{\left[i(n^{2}-2nN_{n})x_{e}\omega_{e}T_{per}\right]}=1.
\end{equation} $ N_{n} $ can be expressed as $ n'+u/v $, where $ n' $
is the integer part and $ u/v $ is the irreducible fraction. Using
this along with Eq. \ref{eq26} and Eq. \ref{eq27}, Eq. \ref{eq30} can
be expressed as
\begin{equation}
\text{exp}{\left(i\left[vn^{2}-2(n'+u)n\right]\dfrac{x_{e}\omega_{e}T_{per}}{v}\right)}=1.
\end{equation} By observing that the square brackets enclose an
integer value, the time period can be calculated as $
T_{per}={2\pi}{x_{e}v/\omega_{e}}$  \cite{strekalov2016}. For
simplicity $ v $ can be taken as unity.

An even Perelomov coherent state is the addition of two Perelomov coherent states with parameters $ \alpha $ and $ -\alpha $, also we are taking $ n' $ to be an even number. Hence the initial wave function is expressed as
\begin{equation}
	\chi(\zeta)=\sum_{m=0}^{n'/2}d_{2m}\Psi_{2m}^{\lambda}(\zeta).
\end{equation}
In Eq. (26), n will be replaced by 2n. Therefore, the time period will be $ T_{per}/4$. For any even higher order superposition with $\ell$ terms, is defined such that the parameter $\alpha$ is multiplied with $\varepsilon_{j}^{(\ell)}$, where $j=0,\,1,\,\dots,\,\ell-1$. The revival time in such a case is $ {T_{per}}/{2\ell}$. Figure \ref{fig2} shows the plots of $\langle x \rangle$ versus $\langle p \rangle$ for the coherent state and even coherent state. The closed figures show the revival of these states. In the above two sections we have seen that when there is a  superposition of wavepackets, the time period of dynamics of the initial states changes. It is a clear indication that the  dynamics  of quantum states not only depends on the Hamiltonian but also on the initial states considered. In the next section, we will discuss other types of dynamics which can occur in quantum
system using time series analysis.

\section{Quasi-periodic, ergodic and chaotic dynamics}\label{sec3}

Nonlinear time series analysis is being widely used as a tool to study
the complicated dynamics of systems using a series of data points
listed in time order. It is highly useful for the understanding of
many complex phenomena in nature. There exist several methods to
compute dynamical parameters such as information dimension, entropy,
Lyapunov exponents, etc. from time series analysis \cite{eckmann1981}.
In this section, we will use some of these methods such as first
return time distribution, recurrence plot and Lyapunov exponent  to
study the dynamics of superposition states.\\

\emph{1. First return time distribution }($ F_{1} $)\\
The first return time can be used to analyze the various dynamical
properties of complex systems. Extensive use of this can be seen in
the literature, for example see \cite{hirata1995}. $ F_{1} $
distribution contains information about the recurrence of a small range
of values in a large time series. Cells of suitable size are
constructed to calculate the frequency of recurrence of data points
within the cell. It was shown that for systems having ergodic dynamics
the $ F_{1} $ distribution can be very well fitted by the exponential
distribution $\mu e^{-\mu\tau}$ \cite{hirata1993,balakrishnan2001}
where $ \mu $ is related to the mean recurrence time $
\langle\tau\rangle $ as $ \mu=\langle\tau\rangle^{-1}$ which follows
from Poincar\'e recurrence theorem.\\

\emph{2. Recurrence plot }\\
Recurrence plots are a recent method for the analysis of nonlinear
data. It was introduced in the famous paper of Eckmann, Kamphorst and
Ruelle \cite{eckmann1987} as a new tool which could extract more
information that is not easily obtained by other methods. In other
words, recurrence plot provide a simple way to visualize the
trajectory in phase space. Our phase space is of higher dimension,
hence cannot be pictured. Recurrence plot helps us to get certain
information about this higher dimensional phase space through a two
dimensional representation and also it depicts the pair of time at
which the trajectory is at the
same point or the point which is sufficiently close(within an $
\epsilon $ neighborhood).
Hence recurrence can be represented by the function
\begin{equation}
R(i,j)=
\begin{cases}
1 & \quad \text{if} \,\left|\left|x(i)-x(j)\right|\right|\leq\epsilon\\
0 & \quad \text{otherwise,}
\end{cases}
\end{equation}
where $x(\cdot)$ is the location of the trajectory and $(i,j)$ are
coordinate points \cite{zbilut1992}. In the 2006 paper of Marwan
\cite{marwan2007}, basic idea of recurrence plot, recurrence
quantification analysis and its applications in various fields are
discussed. Mostly in recurrence plot, parallel, equidistant diagonal
lines indicates periodic trajectories, more than one set of parallel,
diagonal lines or carpet like patterned structure gives
quasi-periodicity and a single diagonal line which may or may not be
surrounded by short broken lines at random distances from this line
shows chaotic trajectories.\\

\emph{3. Lyapunov exponent }\\
Lyapunov exponent ($ \lambda $) is a quantitative measure of the
exponential measure caused due to small change in initial conditions.
In the chaotic regime if the initial separation of two orbits is $
s(0) $, then at later time $ t $ their separation is given by $
s(t)=s(0)e^{\lambda t} $ where $ \lambda $ is a positive number. Here
we have  estimated the maximal Lyapunov exponent ($ \lambda_{max} $)
from the time series using the algorithms developed by Rosenstein et
al. \cite{rosenstein1993} and Kantz \cite{kantz1994} and also verified
that our results stands by repeating our calculations using the
procedure by Wolf et al. \cite{wolf1985}. We can compute the maximum
lyapunov exponent from the plot of $ S(\epsilon,m,t) $ vs t. Here
\begin{equation}
S(\epsilon,m,t)=\Bigg <
ln\Big(\frac{1}{\modu{U_{n}}}\sum_{s_{n'}\epsilon
U_{n}}\modu{s_{n+t}-s_{n'+t}}\Big) \Bigg>_{n}
\end{equation} where $ s_{n'} $ is a very close return to a previously
visited point $ s_{n} $ in the embedding space and $ U_{n} $ is the
superset of all such $ s_{n'} $.
If $ S(\epsilon,m,t) $ exhibits a linear increase with identical slope
for all m larger than some $ m_{0} $ and for a reasonable range
of $ \epsilon $, then this slope can be taken as an estimate of the
maximal exponent.

%%%%%%%%%%%%%%%

\subsection{Coherent state and its superposition in a Kerr medium with cubic nonlinearity}
Let us consider a Hamiltonian for a single-mode electromagnetic field interacting with the atoms of a nonlinear medium,
	\begin{equation}
	H_{1}=\hslash(\chi {\hat{a}^{\dagger 2}}\hat{a}^{2}+\chi'{\hat{a}^{\dagger 3}}\hat{a}^{3}),
	\end{equation}
where $ \hat{a} $ and $ \hat{a}^{\dagger} $ are the photon annihilation and creation operators which satisfy $[\hat{a},\hat{a}^{\dagger}]=1$. The first term in the Hamiltonian models a Kerr medium with a coupling strength $\chi$ and the second term is the one with cubic nonlinearity with a strength $\chi'$. Because of the presence of this nonlinear term the relatively simple periodic behavior of the system is lost. Both ${\hat{a}^{\dagger 2}}\hat{a}^{2}$ and ${\hat{a}^{\dagger 3}}\hat{a}^{3}$ are diagonal in the number operator $N= \hat{a}^{\dagger}\hat{a}$. Hence  ${\hat{a}^{\dagger 2}}\hat{a}^{2}$ can be written as $N(N-1)$ and ${\hat{a}^{\dagger 3}}\hat{a}^{3}$ can be written as $N(N-1)(N-2)$. 	For a generic initial wave packet when $\chi'\neq 0$ exact revivals do not occur and in the space of observable, periodic returns of observables to their initial value is replaced by quasi-periodicity \cite{sudheesh2010}.

%	\begin{figure*}
%%	\centering
%	\includegraphics[scale=0.6]{fig3.eps}
%	\caption{ $ F_{1} $ distribution of the expectation value $ \langle x \rangle. $ Initial state $ \ket{\alpha} $ with $ \modu{\alpha}^{2}=4. $ and $ \modu{\alpha}^{2}=25. $ The ratio of $ \chi'/\chi $ is taken as 0.01.}	\label{fig3}
%	\end{figure*}
%	\begin{figure*}
%%	\centering
%	\includegraphics[scale=0.6]{fig4.eps}
%	\caption{$ F_{1} $ distribution of the expectation value $ \langle x \rangle. $ Initial state $ \ket{\alpha} $ with $ \modu{\alpha}^{2}=9. $ The ratio of $ \chi'/\chi $ is taken as 0.1 and 0.6 respectively.}\label{fig4}
%	\end{figure*}
Let our initial state be an ordinary coherent state $ \ket{\alpha(0)}$ which is the same as in Eq. (4). After time evolution under the Hamiltonian $ H_{1}$, this becomes $ \ket{\alpha(t)}$
\begin{equation}
	\ket{\alpha(t)}=e^{-{\modu{\alpha}^{2}}/{2}}\sum_{n=0}^{\infty}\frac{\alpha^{n}}{\sqrt{n!}}\left( e^{-i\chi n(n-1)t-i\chi'n(n-1)(n-2)t} \right)\ket{n}.
\end{equation}
%For ordinary coherent state  the expectation value of the quadrature $ x=(a+a^{\dagger})/\sqrt{2}$ is
%\begin{eqnarray}
%&&\expect{\psi(t)}{x}{\psi(t)}=e^{-\modu{\alpha}^{2}}\Bigg[\sum_{n=1}^{\infty}\frac{\modu{\alpha}^{2n}\alpha^{*-1}}{(n-1)!}\Big( e^{-i\chi (n-1)2t}\nonumber\\
%&&+ e^{-i\chi'(n-1)(n-2)3t}+e^{i(n-1)((n-2)(n-3)\chi'-n\chi)t}\nonumber\\
%&&+e^{i(n-1)(n-2)(\chi-n\chi')t}\Big)+\sum_{n=0}^{\infty}\frac{\modu{\alpha}^{2n}\alpha^{*}}{n!}\Big( e^{i\chi n2t}+ e^{i\chi'n(n-1)3t}\nonumber\\
%&&+e^{in(n-1)((n+1)\chi'-\chi)t}+e^{in((n-1)\chi-(n-1)(n-2)\chi')t}\Big)\Bigg].
%\end{eqnarray}

The analysis of $F_{1}$ distribution and recurrence plot with different $\modu{\alpha}^{2}$ value by keeping  $\chi'/\chi$ ratio fixed and vice-versa has already been carried out \cite{sudheesh2009recurrence,sudheesh2010}. They have shown the appearance of hyperbolicity and ergodicity in the dynamics of the system.\\
As in Sec. \ref{sec2}, we expect a change in dynamics when superposition of states are considered. Here we are comparing the $F_{1} $ distribution and recurrence plot of the expectation value of $\langle x^{2} \rangle $ for the states with $ \ell=1 $ and $2$ (Eq. 9). For ordinary coherent state $ \aver{x^{2}} $ comes out to be
\begin{equation}
\begin{aligned}
\aver{x^{2}}={} & \dfrac{1}{2}+\modu{\alpha}^{2}+e^{-\modu{\alpha}^{2}}\Big(\sum_{n=0}^{\infty}\frac{\modu{\alpha}^{2n}\alpha^{2}}{n!}e^{i\big(2(2n+1)\chi+6n^{2}\chi'\big)t}\\& + H.c.\Big).
\end{aligned}
\end{equation}

\begin{figure}[h]
\centering
\subfigure[]{\includegraphics[width=0.48\linewidth]{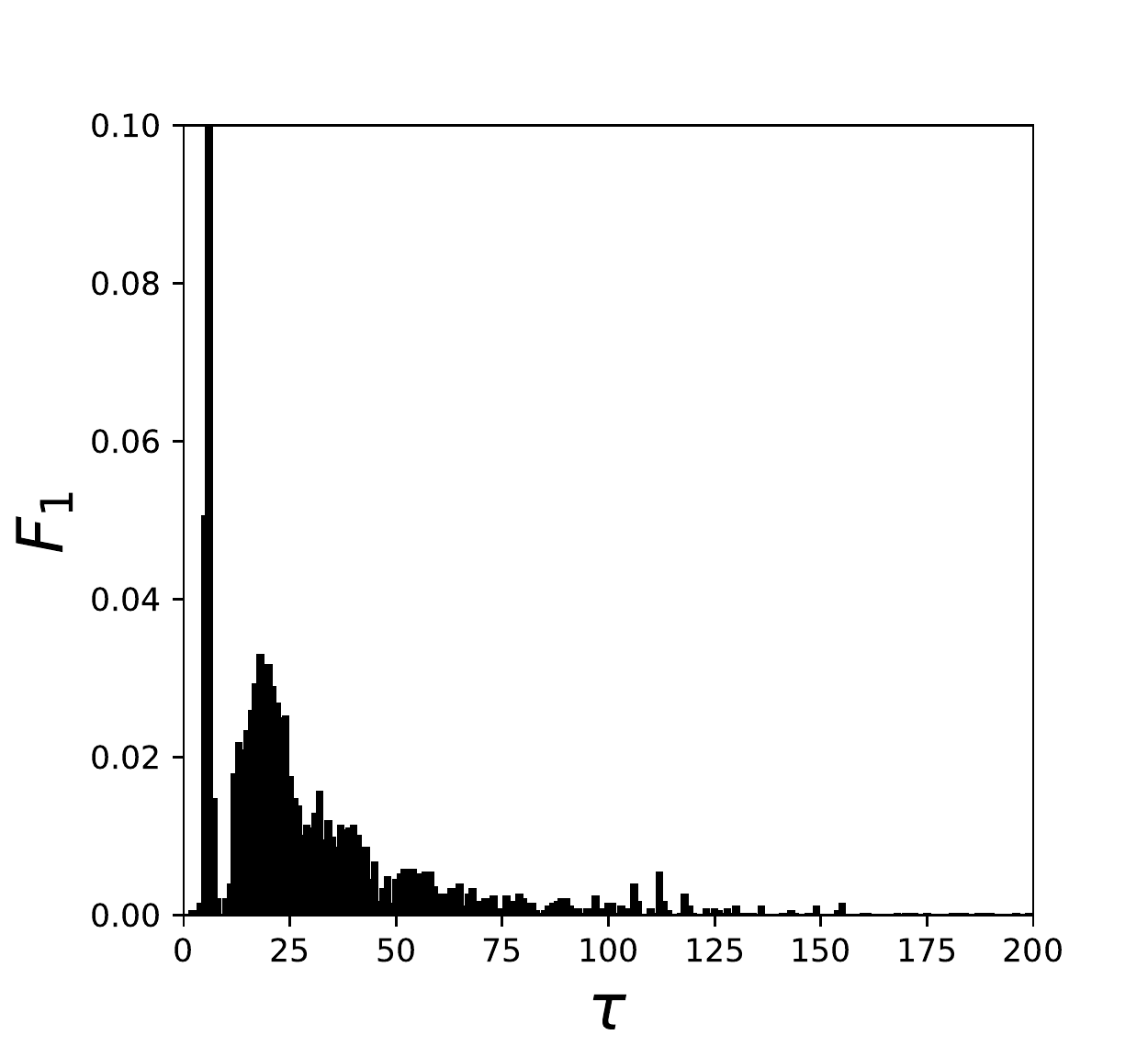}}
\hspace{-.01\linewidth}
\subfigure[]{\includegraphics[width=0.49\linewidth]{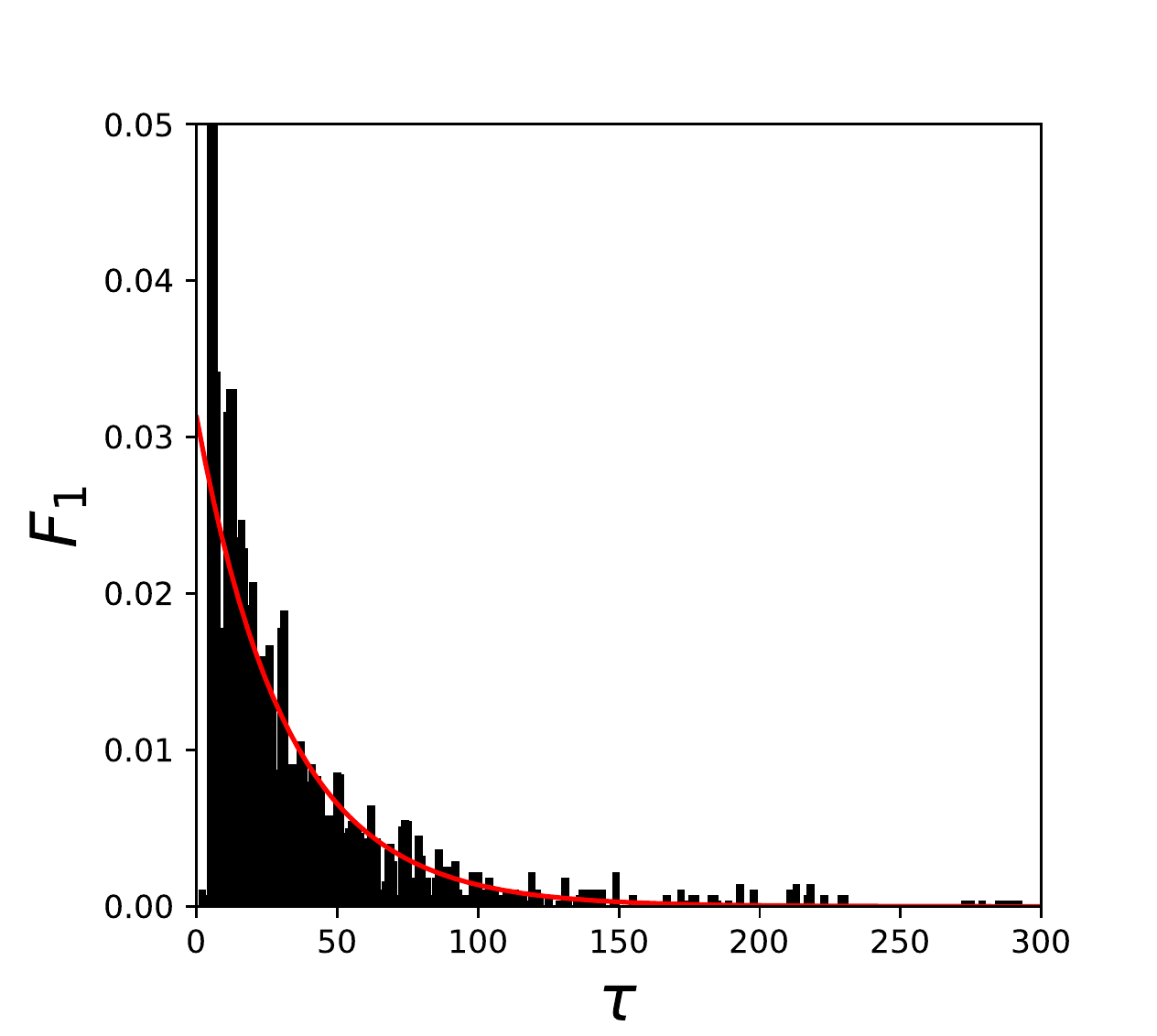}}
\caption{$ F_{1} $ distribution with $ \modu{\alpha}^{2}=25 $ and $
\chi'/\chi=10^{-3} $ for (a) coherent state and (b) even coherent state.}\label{fig3}
\centering
\end{figure}

\begin{figure}[h]
\centering
\subfigure[]{\includegraphics[width=0.49\linewidth]{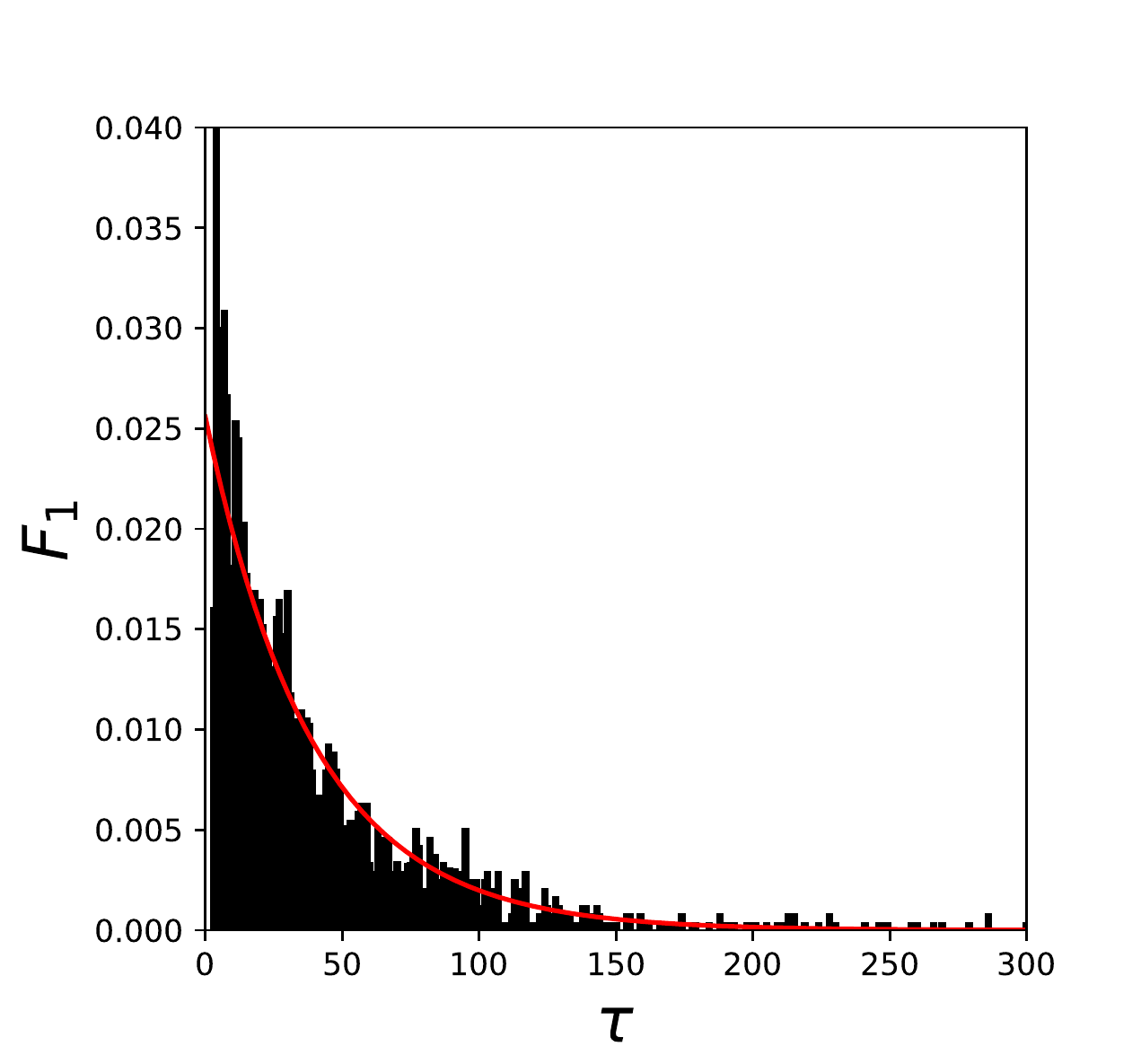}}
\hspace{-.01\linewidth}
\subfigure[]{\includegraphics[width=0.49\linewidth]{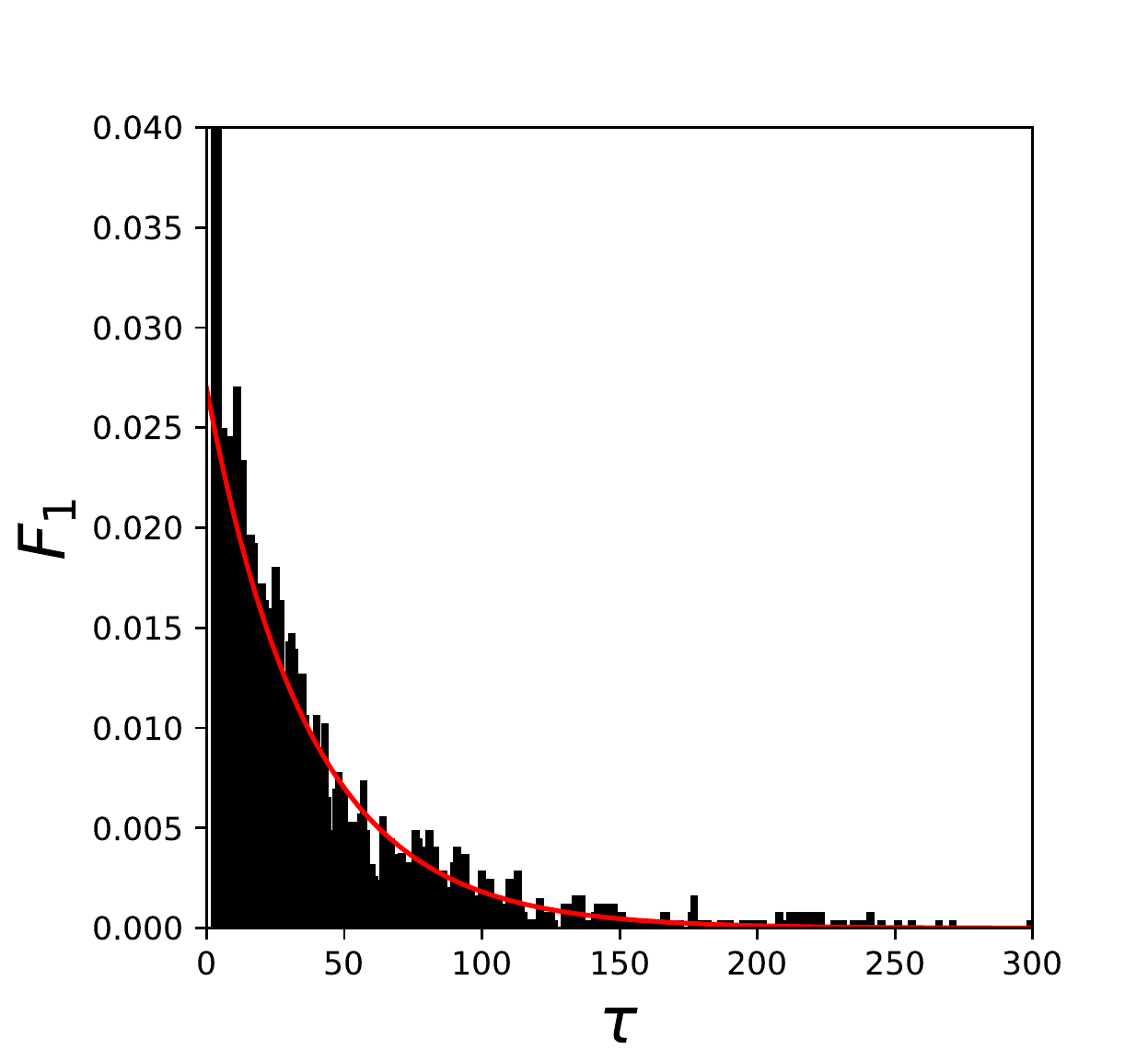}}
\caption{$ F_{1} $ distribution with $ \modu{\alpha}^{2}=100 $, $
\chi'/\chi=10^{-3} $ for (a) coherent state and (b) even coherent state.}\label{fig4}
\end{figure}

%\begin{figure}[h]
	%\centering
	%\begin{minipage}{.46\linewidth}
	%	\includegraphics[width=\linewidth]{fig3a.pdf}
	%\end{minipage}
	%\hspace{.05\linewidth}
	%\begin{minipage}{.46\linewidth}
	%	\includegraphics[width=\linewidth]{fig3b.pdf}
	%\end{minipage}
	%\caption{$ F_{1} $ distribution with $ \modu{\alpha}^{2}=25 $, $ \chi'/\chi=10^{-3} $ for CS and even coherent state.}\label{fig3}
	%\centering
	%\begin{minipage}{.46\linewidth}
	%	\includegraphics[width=\linewidth]{fig4a.pdf}
	%\end{minipage}
	%\hspace{.05\linewidth}
	%\begin{minipage}{.46\linewidth}
%		\includegraphics[width=\linewidth]{fig4b.pdf}
	%\end{minipage}
	%\caption{$ F_{1} $ distribution with $ \modu{\alpha}^{2}=100 $, $ \chi'/\chi=10^{-3} $ for CS and even coherent state.}\label{fig4}
%\end{figure}

\begin{figure}[h]
\centering
\subfigure[]{\includegraphics[width=0.49\linewidth]{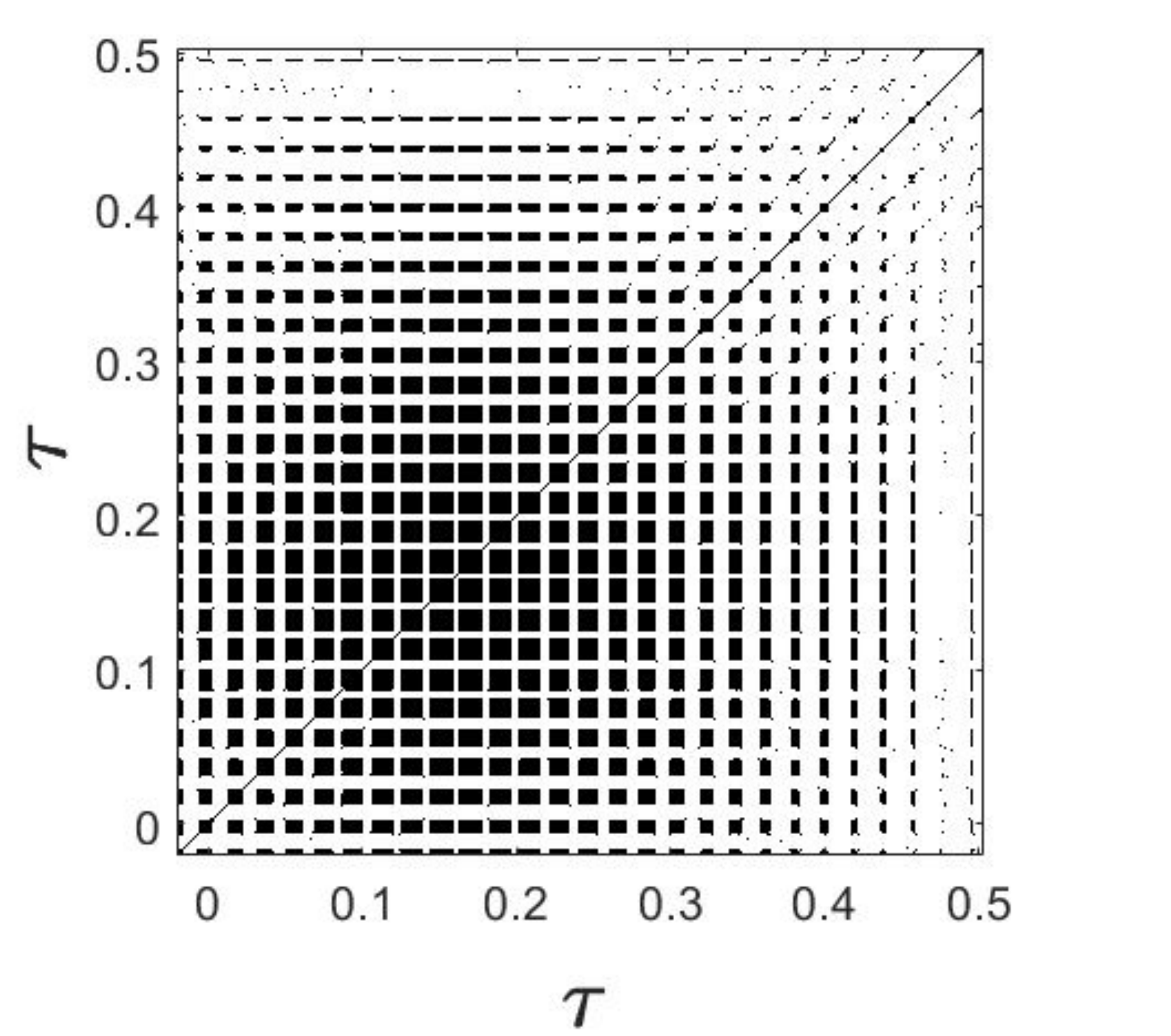}}
\hspace{-.01\linewidth}
\subfigure[]{\includegraphics[width=0.45\linewidth]{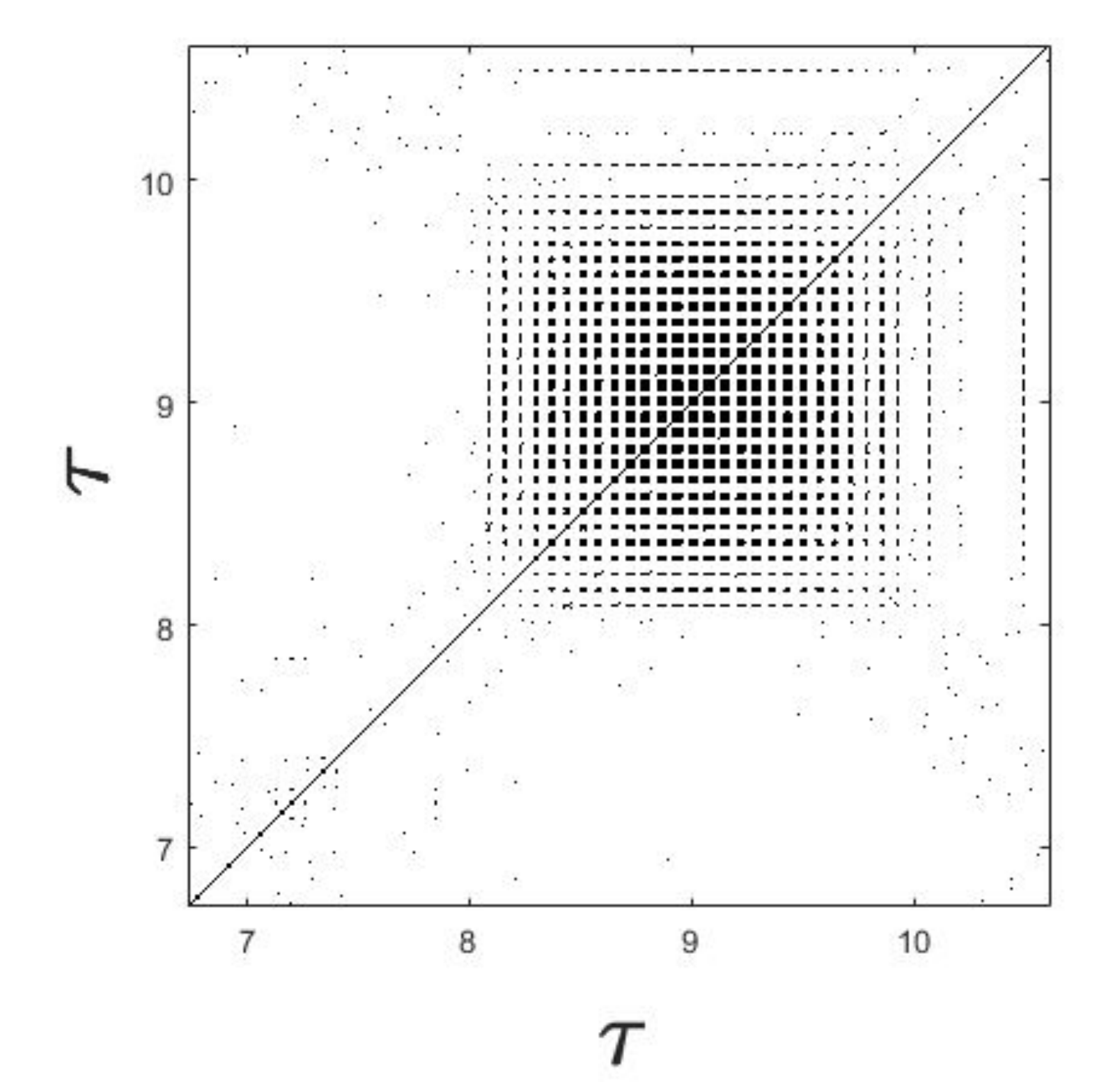}}
\caption{Recurrence plot
  for initial (a) coherent state and (b) even coherent state with $ \modu{\alpha}^{2}=25 $ and $
 \chi'/\chi=10^{-3} $ (the parameters are same as in Fig. 3).
}\label{fig5}
\centering
\end{figure}

\begin{figure}
\centering
\subfigure[]{\includegraphics[width=0.49\linewidth]{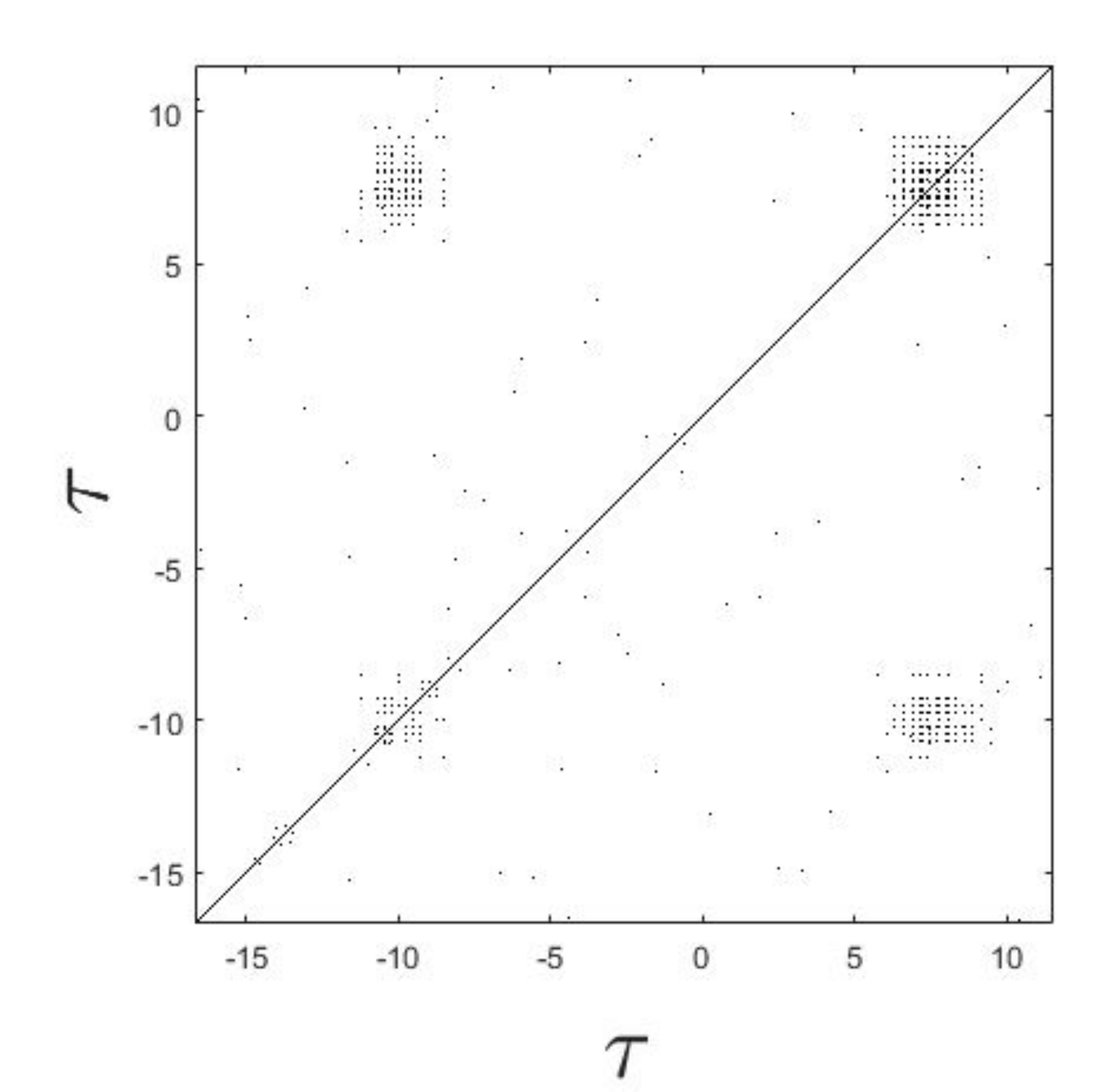}}
\hspace{-.01\linewidth}
\subfigure[]{\includegraphics[width=0.49\linewidth]{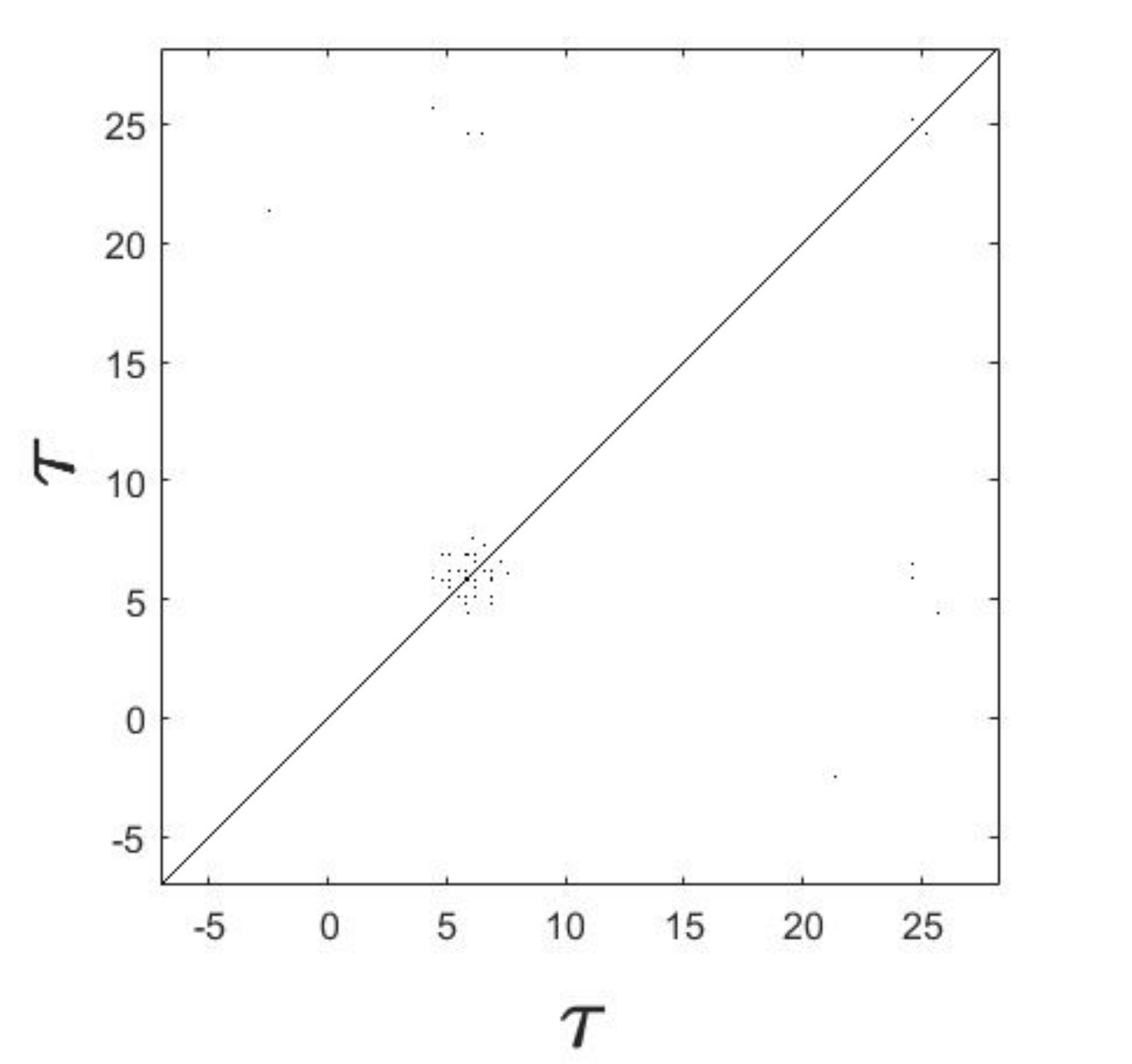}}
\caption{Recurrence plot  for  initial (a) coherent state and (b) even coherent state  with $ \modu{\alpha}^{2}=100 $, $
\chi'/\chi=10^{-3}$ (parameters are same as in Fig. 4). }\label{fig6}
\end{figure}

Similarly, we have computed this quantity for even  ($ \ell=2 $) and higher order superposition state. Fig. \ref{fig3} compares the $ F_{1} $ distribution for coherent state and even coherent state (Eq. 13)  for $ \modu{\alpha}^{2}=25 $ and $ {\chi'}/{\chi}=10^{-3} $ with $ 10^{6} $ data points and cell size between $ 10^{-3}$ and $ 10^{-2} $. The $ F_{1} $ distribution for CS $ \ket{\alpha} $, for small $ \modu{\alpha}^{2}  $ is a discrete one with finite number of points which shows the quasi-periodic nature. If, instead we use even coherent state a decaying exponential distribution is obtained, which signifies a higher degree of mixing and  the presence of ergodicity in the dynamics. This is more pronounced for larger value of $ \modu{\alpha}^{2}$, as may be seen in Fig. \ref{fig4}. Figures \ref{fig5} and  \ref{fig6} depicts the recurrence plots corresponding to the $ F_{1} $ distribution in Figs. \ref{fig3} and \ref{fig4} respectively. The regular patterned structure is a characteristic of quasi-periodicity  with small number of incommensurate frequencies \cite{marwan2007}. The other recurrence plots signals the increase in the degree of mixing with superposition and  with the increase in $ \modu{\alpha}^{2} $, corroborating our deduction based on $ F_{1} $ distribution. Similarly higher order superposition is also analyzed and similar conclusion is drawn (not shown here).
\subsection{Dynamics of superposition states in the bosonic Josephson junction}
With the advent of Bose-Einstein condensate (BEC) of weakly interacting gases \cite{anderson1995observation, davis1995bose, bradley1995evidence}, an experimental system has become available for the quantitative investigation of Josephson effects in a very well controllable environment. We have focused on bosonic Josephson junction, generated by confining single BEC in a double-well potential. We have considered the Bose-Hubbard Hamiltonian for N bosons in a two-site system.
\begin{equation}\label{HBEC}
H_{2}=-\dfrac{J}{2}({\hat{a}_{1}^{\dagger}}\hat{a}_{2}+\hat{a}_{1}\hat{a}_{2}^{\dagger})+\dfrac{U}{4}(\hat{a}_{1}^{\dagger}\hat{a}_{1}-\hat{a}_{2}^{\dagger}\hat{a}_{2})^{2},
\end{equation}
where $ a_{i} $ and $ a_{i}^{\dagger} $ are the annihilation and creation operators respectively for the bosonic particle in $ i^{th} $ mode. $J$ is the hopping amplitude describing the mobility of bosons and is the measure of coupling strength between the two modes and U, the interaction strength arising from the local interaction within the two wells. By defining the three $ SU(2) $ generators, $ L_{x}=(\hat{a}_{1}^{\dagger}\hat{a}_{2}+\hat{a}_{1}\hat{a}_{2}^{\dagger})/2 $, $ L_{y}=(\hat{a}_{1}^{\dagger}\hat{a}_{2}-\hat{a}_{1}\hat{a}_{2}^{\dagger})/2 $ and $ L_{z}=(\hat{a}_{1}^{\dagger}\hat{a}_{1}-\hat{a}_{2}\hat{a}_{2}^{\dagger})/2 $,  Eq. (\ref{HBEC}) can be expressed in the form \begin{equation}\label{key}
H_{2}=-JL_{x}+UL_{z}^{2}.
\end{equation} Most experiments on the bosonic Josephson system measures quantities defined via the expectation values of single particle Bloch vector such as $\dfrac{ 2\aver{L_{x}}}{N} $. The dimensionless parameter $ u=\dfrac{NU}{J} $ is considered to study the system dynamics. We have taken $ u $ such that it falls in the so-called Josephson regime $ (1<u<N^{2}) $. In the Josephson regime, the fluctuations in the atom numbers are reduced and the coherence is high.

\begin{figure}
\centering
\subfigure[]{\includegraphics[width=0.49\linewidth]{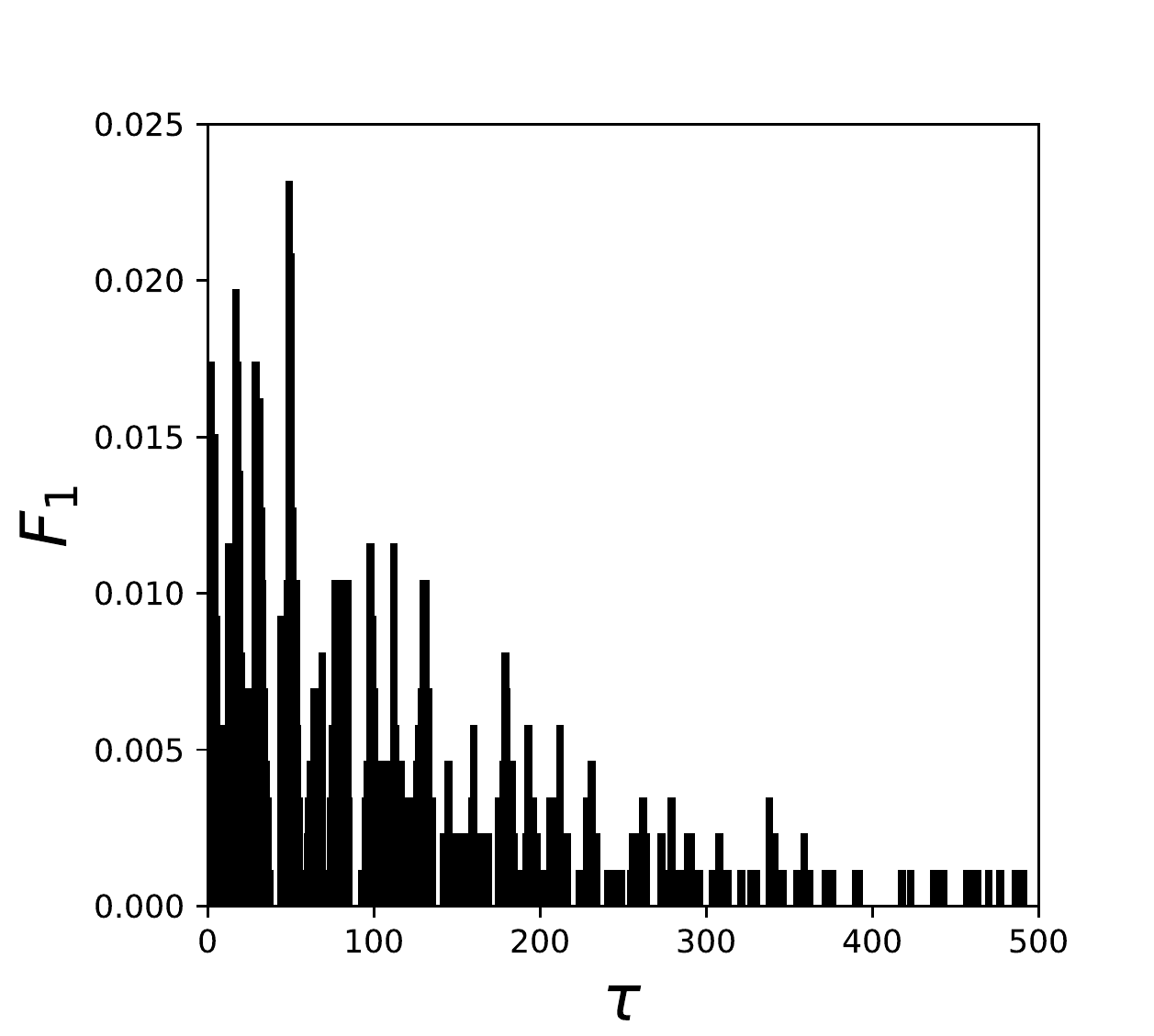}}
\hspace{-.01\linewidth}
\subfigure[]{\includegraphics[width=0.49\linewidth]{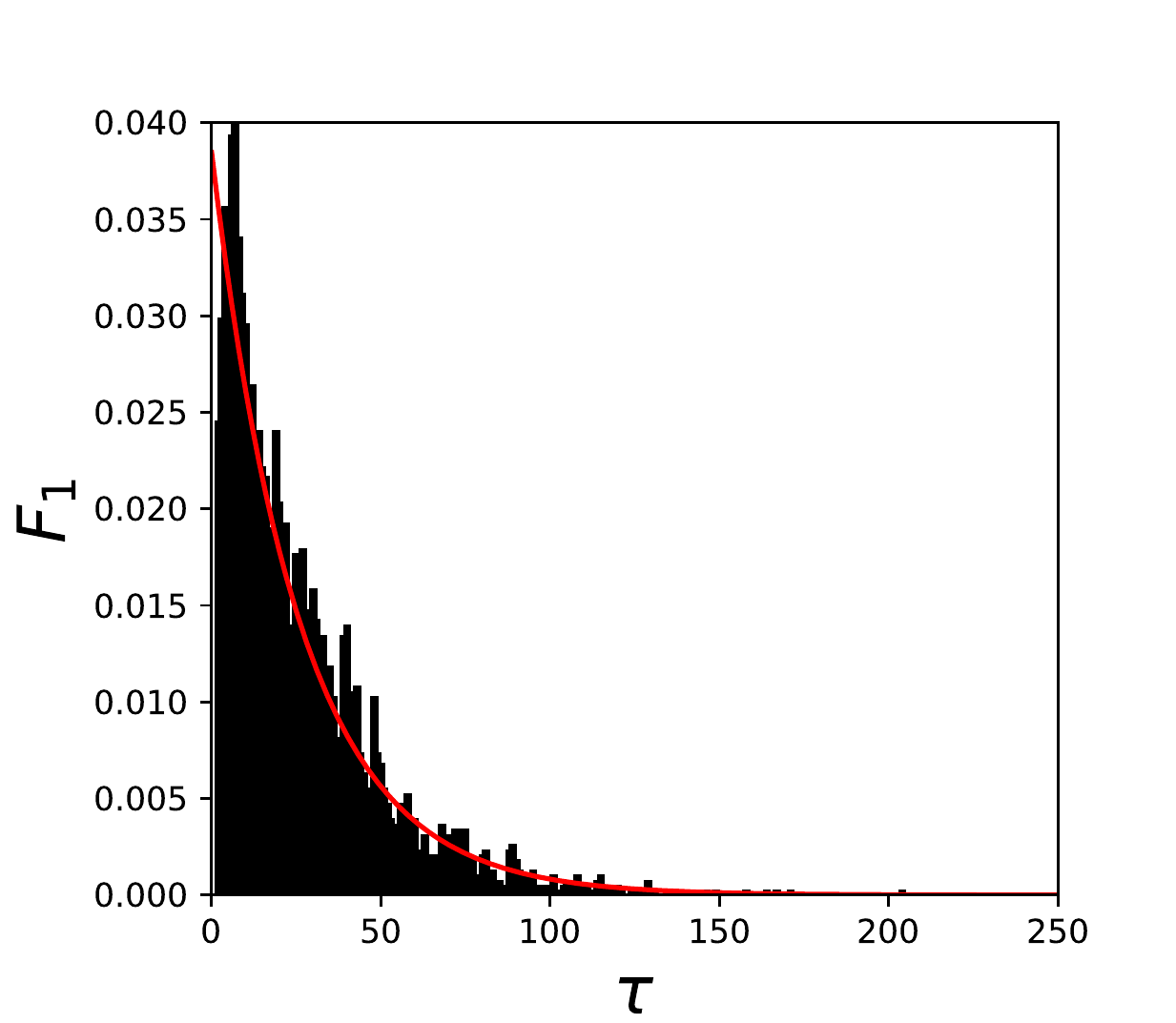}}
\caption{$ F_{1} $ distribution for (a) pi state and (b) even state with $N=40$
and $u=50$.}\label{fig7}
\centering
\end{figure}

\begin{figure}
\vspace{-0.5cm}
\centering
\subfigure[]{\includegraphics[width=0.49\linewidth]{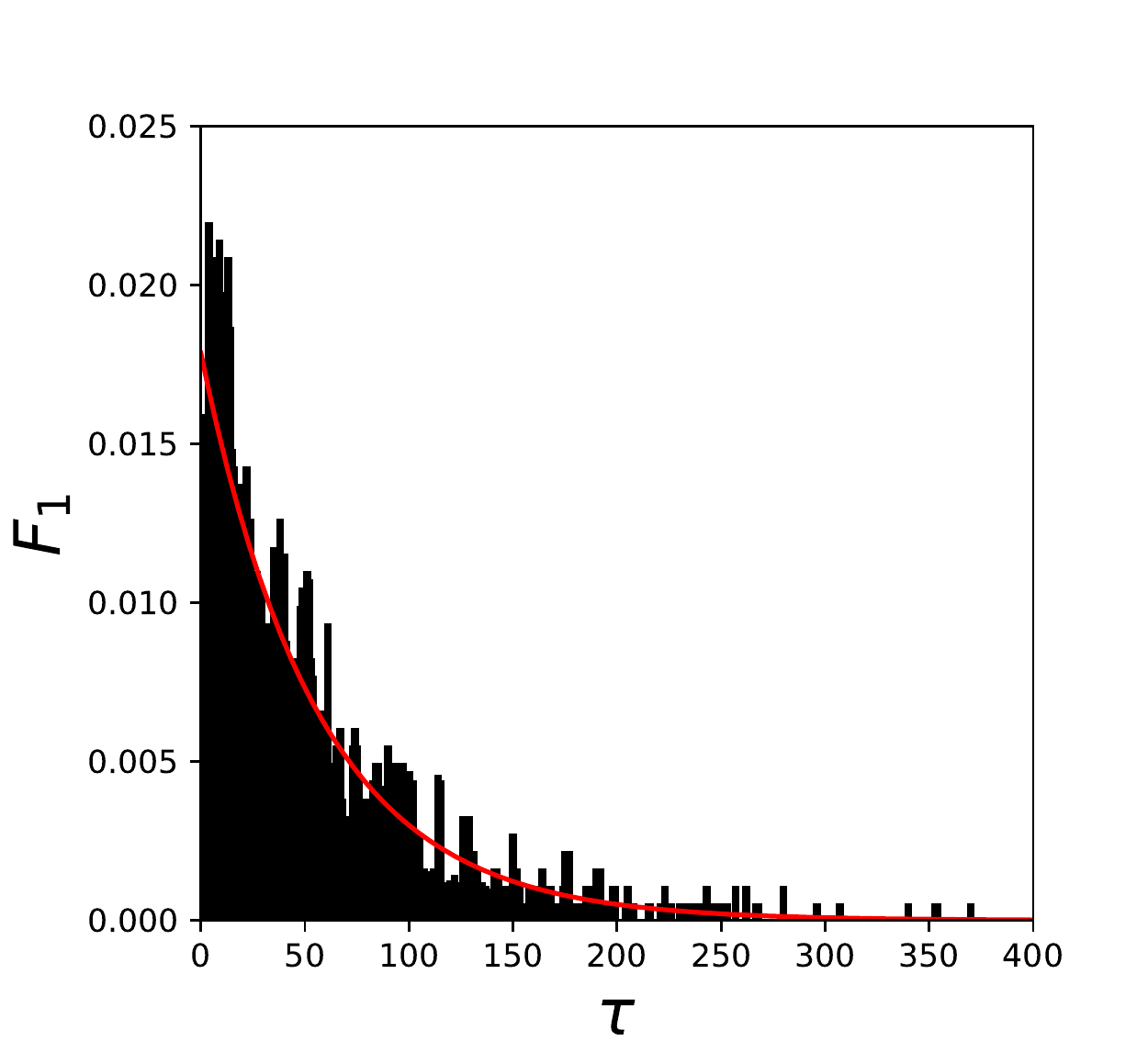}}
\hspace{-.01\linewidth}
\subfigure[]{\includegraphics[width=0.49\linewidth]{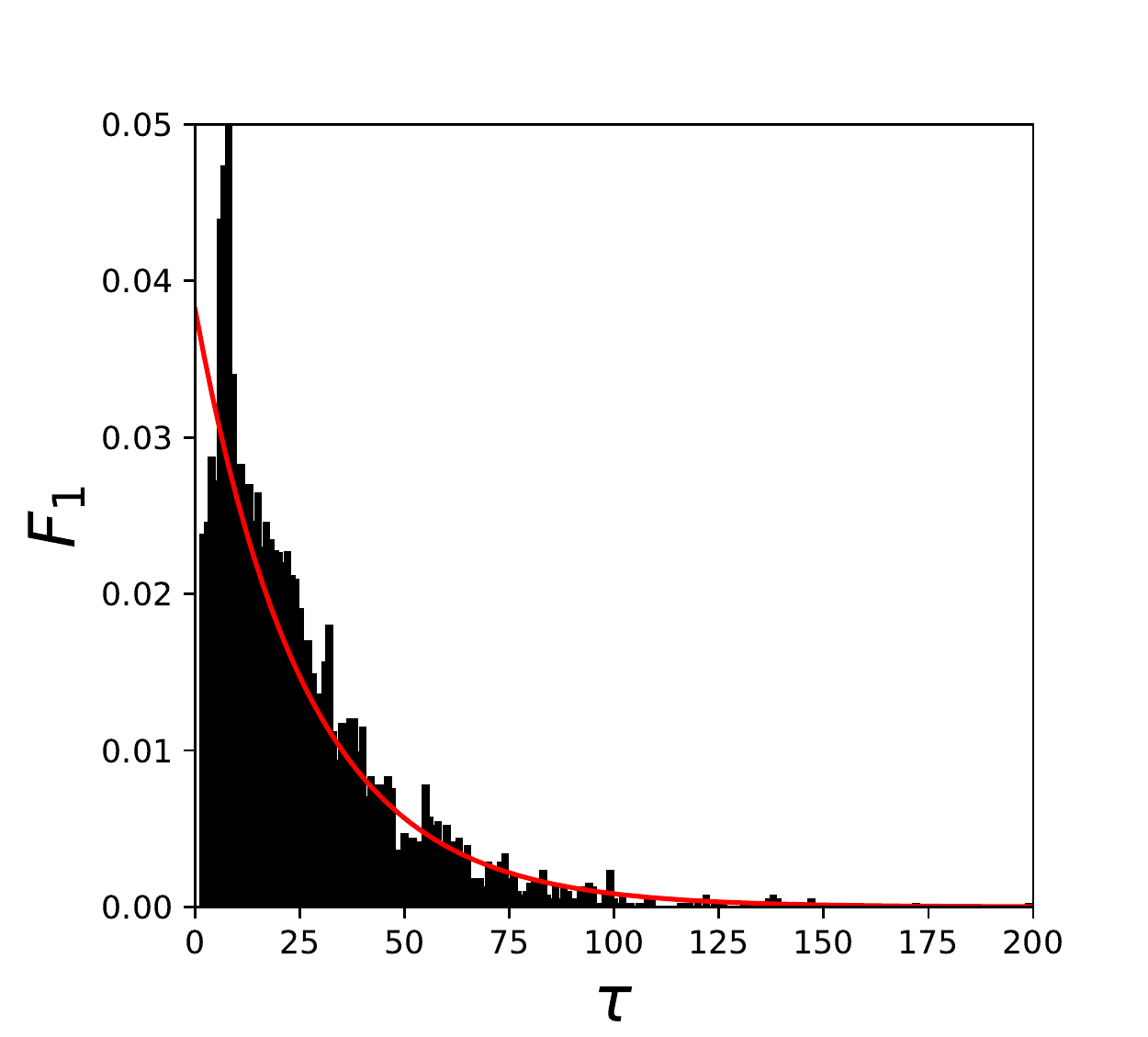}}
\caption{$ F_{1} $ distribution for (a) pi state and (b) even state with $N=40$
and $u=90$.}\label{fig8}
\centering
\end{figure}

\begin{figure}
\centering
\subfigure[]{\includegraphics[width=0.49\linewidth]{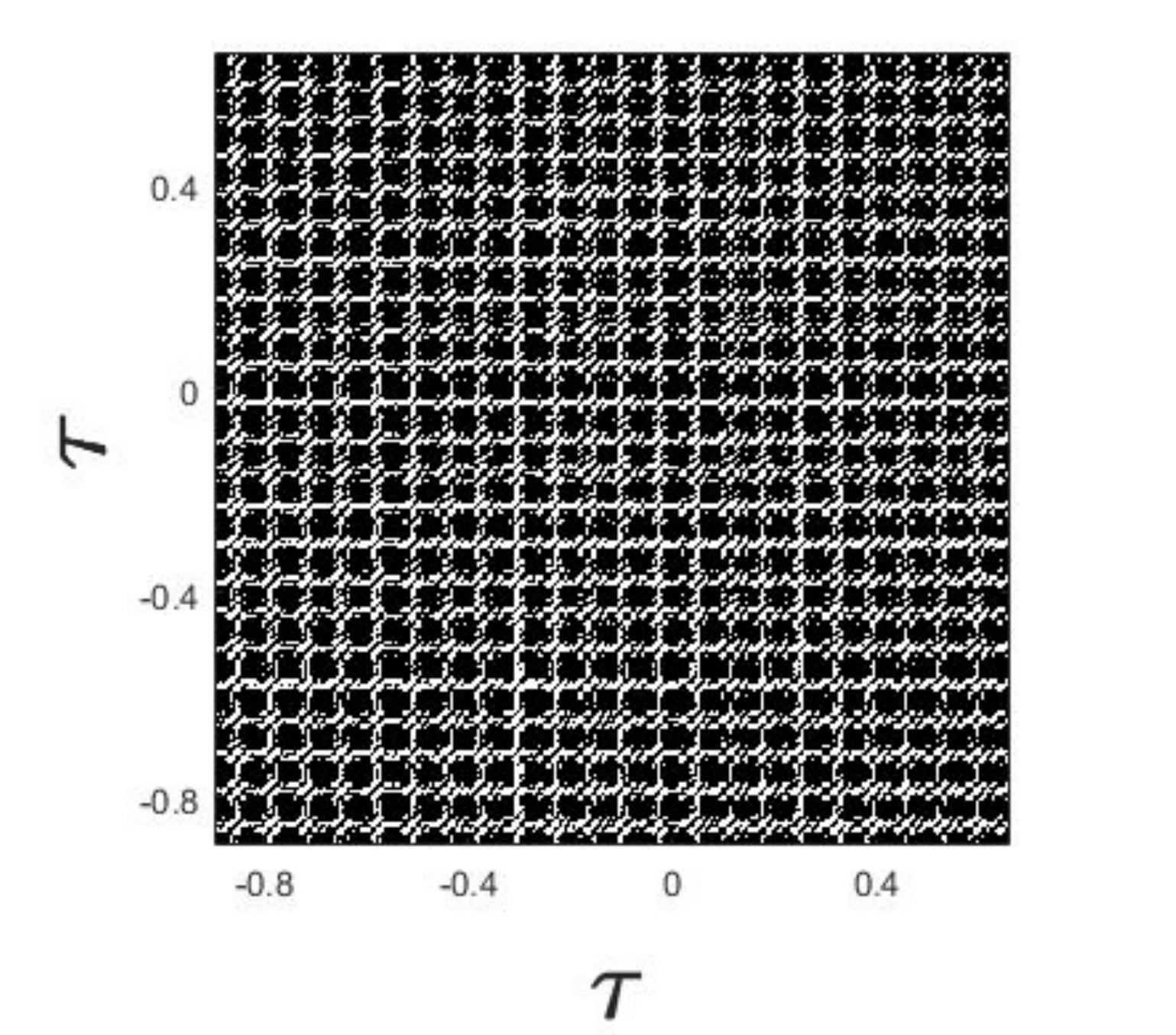}}
\hspace{-.01\linewidth}
\subfigure[]{\includegraphics[width=0.45\linewidth]{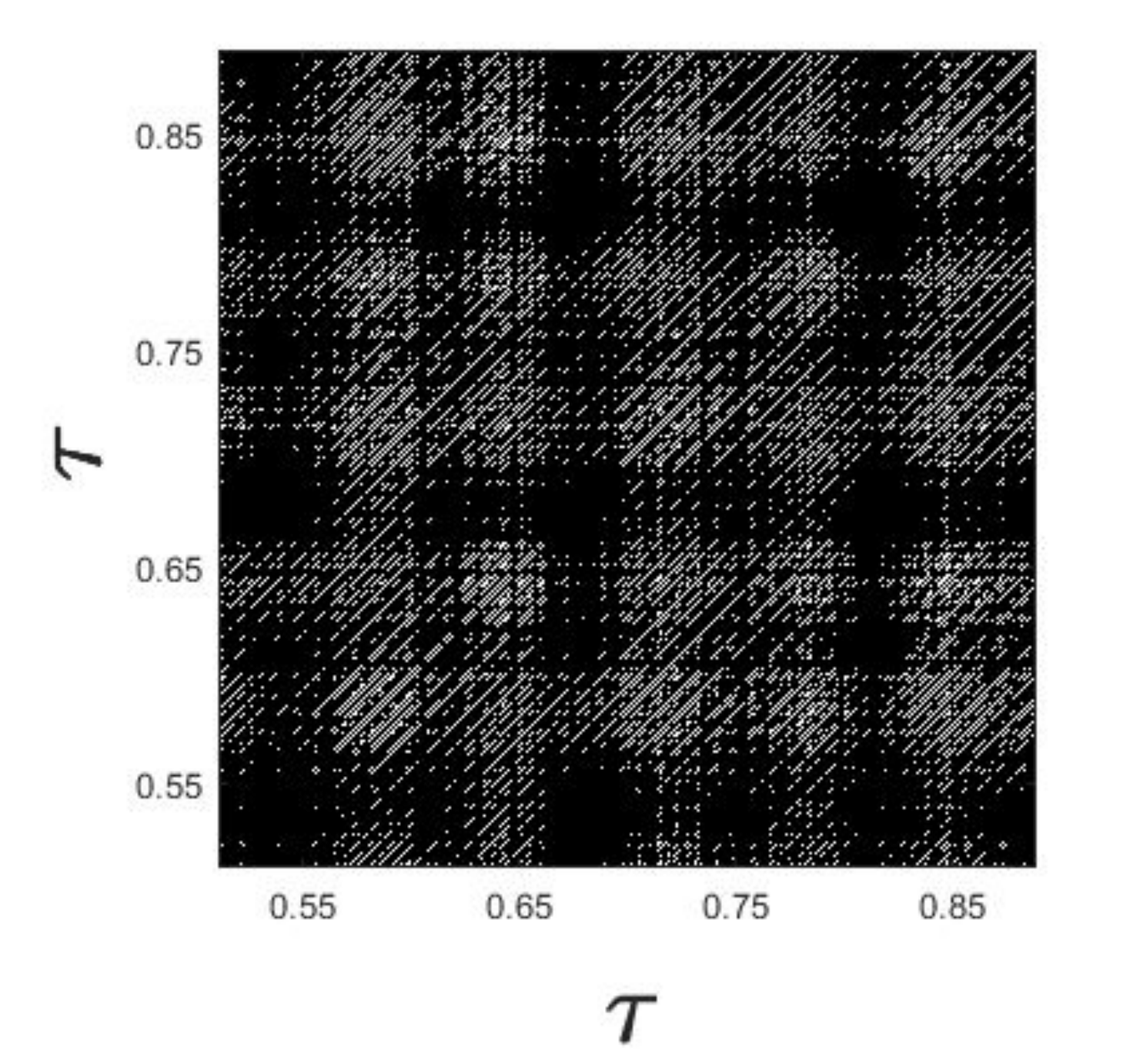}}
\caption{Recurrence plot for initial (a) pi state and (b) even state with $N=40$
and $u=50$ (the parameters are same as in Fig. 7).}\label{fig9}
\centering
\end{figure}

\begin{figure}
\centering
\subfigure[]{\includegraphics[width=0.49\linewidth]{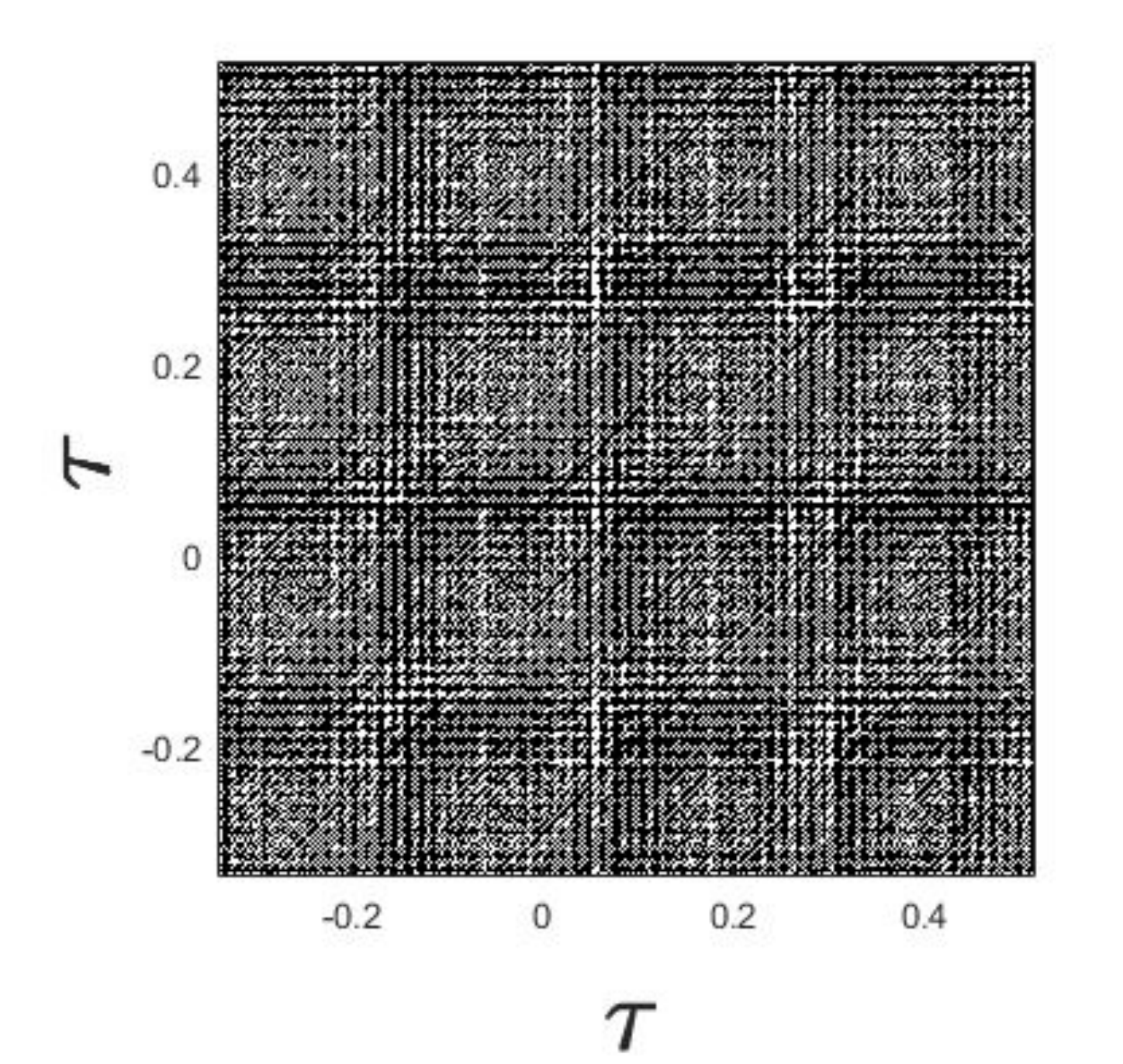}}
\hspace{-.01\linewidth}
\subfigure[]{\includegraphics[width=0.49\linewidth]{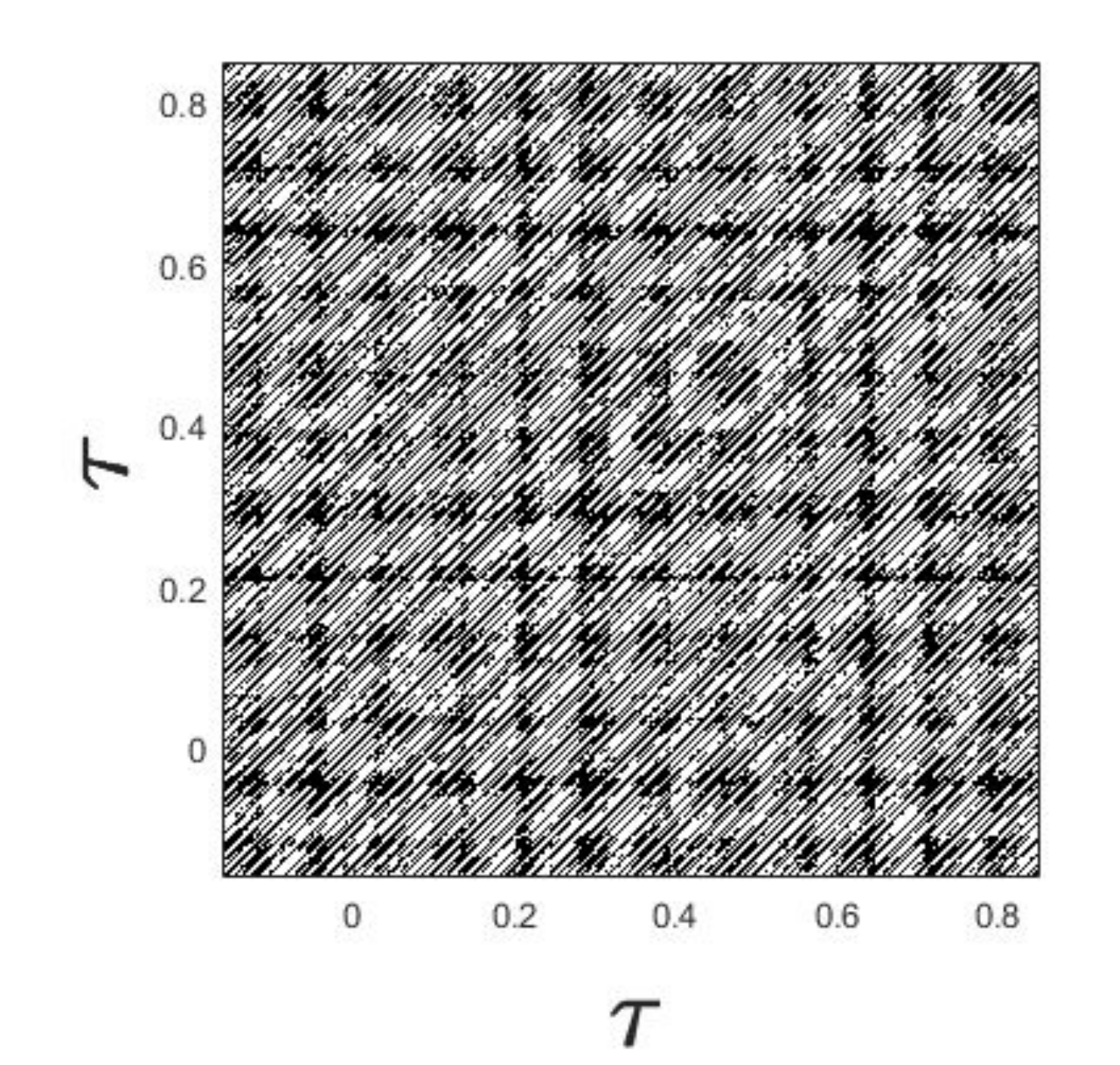}}
\caption{Recurrence plot for initial (a) pi state and (b) even state with $N=40$
and $u=90$ (the parameters are same as in Fig. 8) .}\label{fig10}
\end{figure}

The dynamics of the system is carried out by considering $ SU(2) $ coherent states and its superposition as initial state. $ SU(2) $ coherent states are the eigenstates of angular momentum operator $ {\hat{L}^{2}} $. In terms of angular momentum basis, $ \ket{l;m}\equiv\ket{l+m,l-m} $ where $ m $ varies from $ -l $ to $ l $,  the $ SU(2) $ coherent state can be written as \begin{equation}\label{key}
\begin{aligned}
\ket{\theta,\phi}={} &
\Big[1+\text{tan}^{2}(\theta/2)\Big]^{-1}\sum_{m=-l}^{l}\Big[\text{tan}\left((\theta/2)e^{-i\phi}\right)\Big]^{l+m}\\&\times \binom{2l}{l+m}^{1/2}\ket{l;m},
\end{aligned}
\end{equation} where $ 0<\theta<\pi $ and $ 0<\phi<2\pi $ are the rotation angles of the state $ \ket{m=-l} $. In our study we have taken $ \theta=\pi/2 $, which corresponds to equal population in the two modes.

\begin{figure}
  \centering
  \subfigure[]{\includegraphics[width=0.65\linewidth]{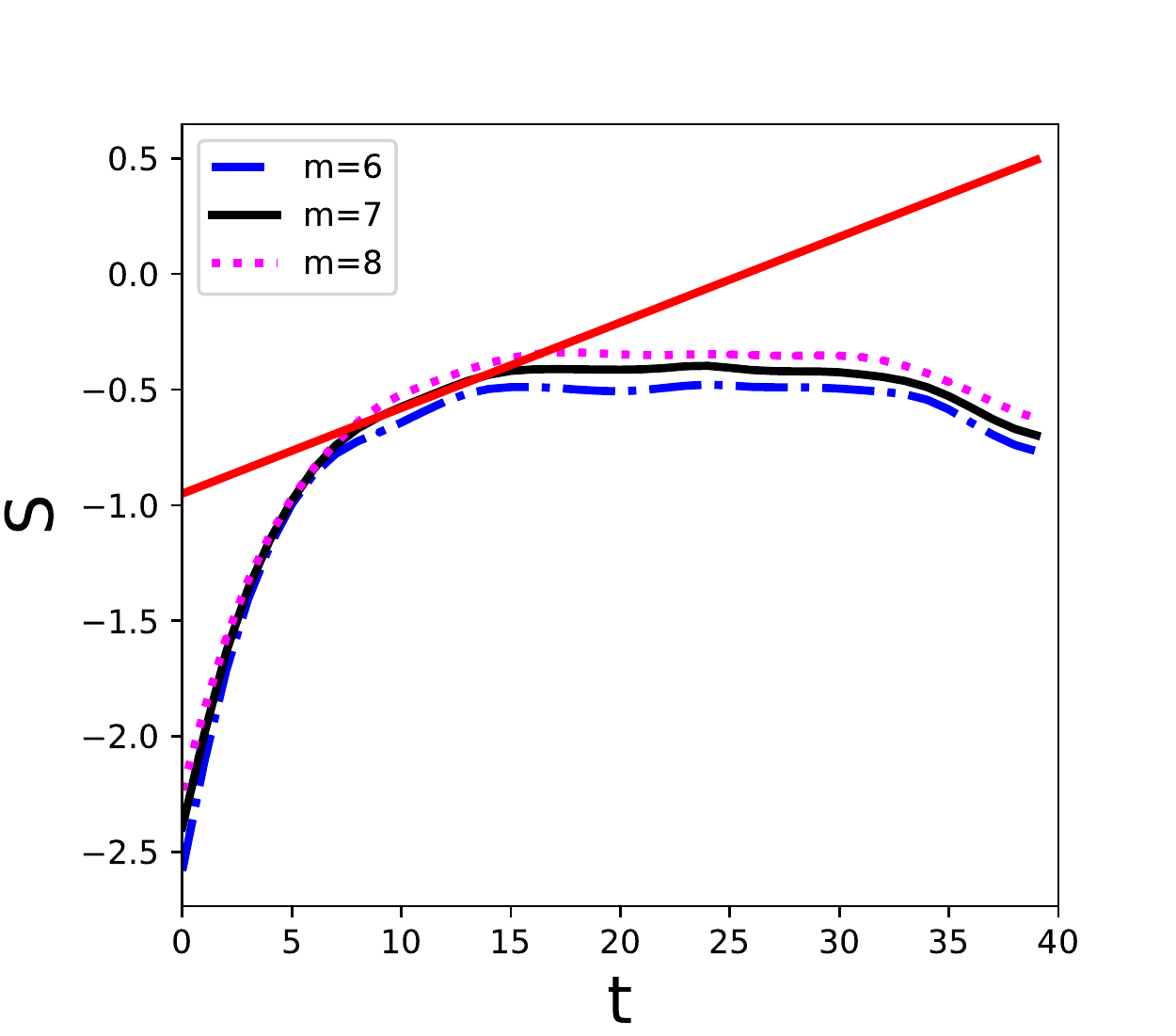}}
\caption{Plot  to find the Lyapunov exponet ($ \lambda$) for initial even state with $N=40$
and $u=50$  (parameters are same as in
Fig. \ref{fig9}). Slope of the graph  gives $\lambda=0.036
$.}\label{fig11}
\end{figure}

\begin{figure}
\centering
\subfigure[]{\includegraphics[width=0.55\linewidth]{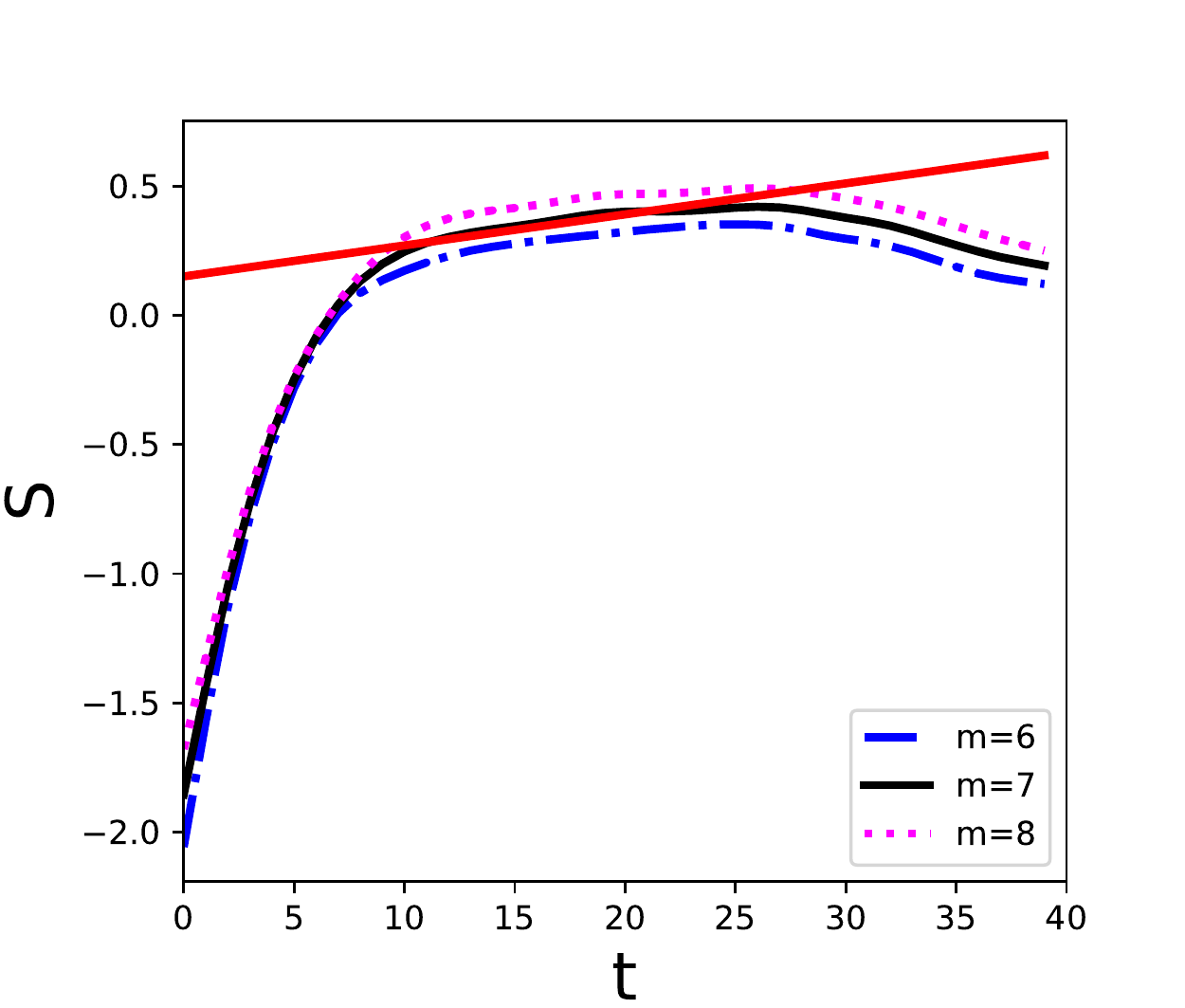}}
\hspace{-.01\linewidth}
\subfigure[]{\includegraphics[width=0.55\linewidth]{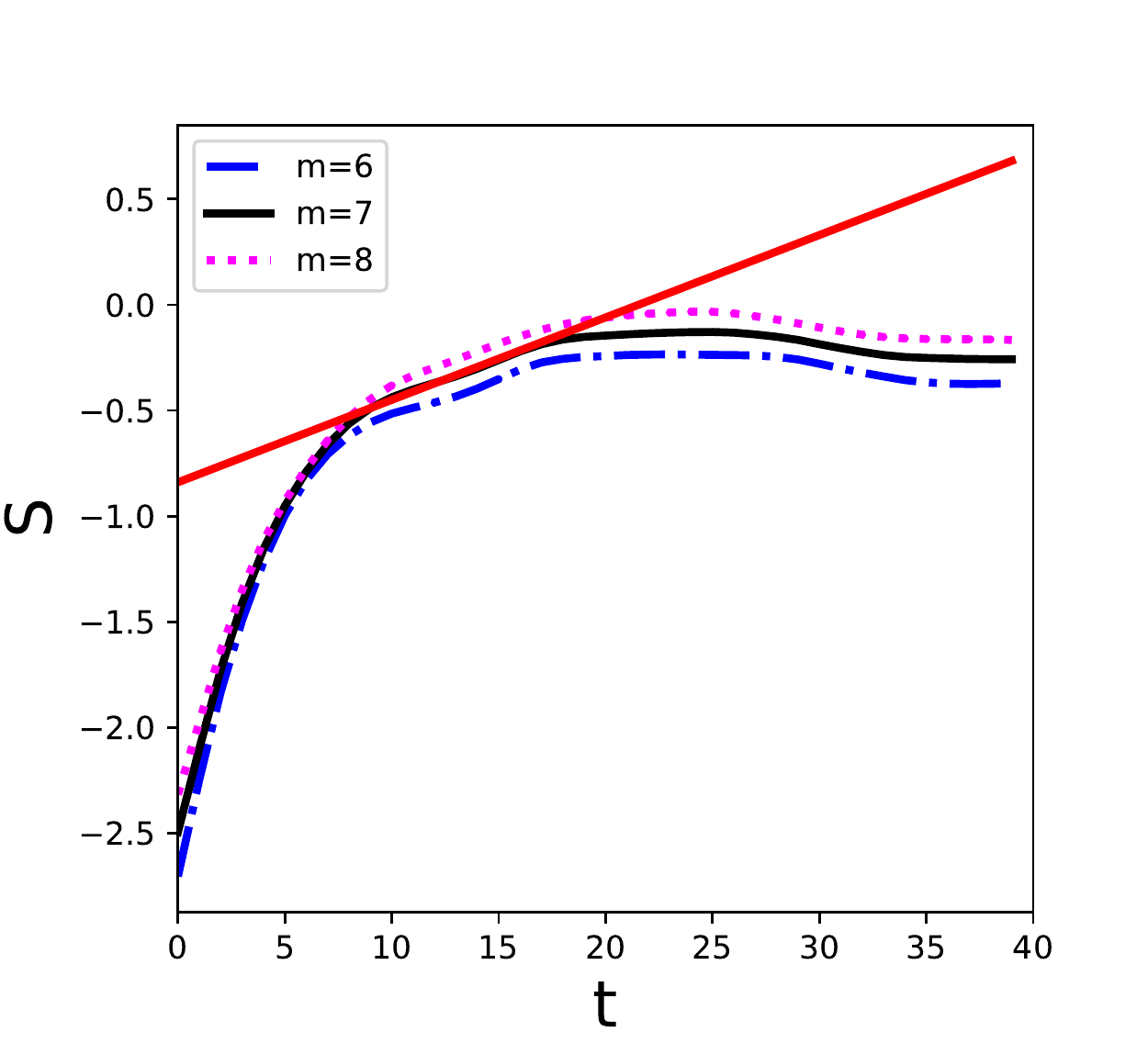}}
\caption{Plots to find the Lyapunov exponent  for initial (a) pi state and (b) even state with $N=40$
and $u=90$ (the parameters are same as in Fig. \ref{fig10}).
Slops of the graphs give  $\lambda=$ (a) $ 0.012$ and  (b) $0.039$.}
\label{fig12}
\end{figure}

 To study and analyze the difference in the dynamics of superposition states, we compare the $ F_{1} $ distribution, recurrence plot and Lyapunov exponent for the states $ \ket{\pi/2,\pi} $ (pi state) and $ \dfrac{1}{\sqrt{2}}\big(\ket{\pi/2,0}+\ket{\pi/2,\pi}\big).$ Depending on whether $ l $ is even or odd the above superposition state can be called even or odd state $ (l=N/2) $.
\\Figs. \ref{fig7}-\ref{fig10} compare the $ F_{1} $ distribution and recurrence plot for pi state and even state with $ N=40 $ and $ 10^{6} $ data points.  Fig. \ref{fig7} depicts the $ F_{1} $ distribution with $ u=50 $ and cell size of $ 10^{-2} $. The quasi-periodicity of the pi state is clear from the finite number of points in the $ F_{1} $  distribution and the hyperbolicity and ergodicity in the dynamics is seen for the superposition (even) state which shows a decaying exponential spectrum. As $ u $ value increases the ergodicity in the dynamics becomes more pronounced as clear from Fig. \ref{fig8}. The regular patterned structure in the recurrence plot in Fig. \ref{fig9} is consistent with the $ F_{1} $ distribution and the broken lines in the other recurrence plot signals the signature of chaos in the system.
When  we consider superposition states,  more number of broken lines are visible in recurrence plots (Fig. \ref{fig10}) which indicates more chaoticity in the system. Lyapunov exponent for these are calculated and chaotic behaviour in the dynamics is confirmed with positive $ \lambda $ and it is seen that $ \lambda $ is larger for superposition states which is evident from Fig. \ref{fig11} and Fig. \ref{fig12}. Other superposition states also gives similar difference in the dynamics from the non-superposed one.

\section{Conclusion}\label{sec4}

We have studied the dynamics of superposition of wavepackets evolving under different nonlinear Hamiltonians corresponding to Kerr medium, Morse oscillator and bosonic Josephson junction. We have found that even the period of evolution changes when we consider different superpositions of states as initial states. Further, we have extended the study to find the consequence of superposition states on    the more complex dynamics such as quasi-periodic, ergodic and chaotic dynamics  using both qualitative and quantitative  methods in time series analysis. We have shown that the systems which are periodic  turned to quasi-periodic or ergodic   when we have changed the initial state from single wave packet to superposition of wave packets. Our results in this paper is a new direction in the theory of nonlinear dynamics in quantum systems because dynamical changes in the evolution of systems due to superposition of wavepackets are not reported in the literature earlier. 

 \section{Acknowledgement}
 M. R. acknowledges support by the Institute for Basic Science in Korea (IBS-R024-Y2).
\bibliography{reference1}
\end{document}